\documentclass[preprint,12pt]{elsarticle}

\usepackage{amssymb}
\usepackage{amsthm}
\usepackage{amsmath}

\usepackage{mathrsfs}
\usepackage{graphicx}
\usepackage{epstopdf}
\usepackage{float}
\usepackage{caption}
\usepackage{subcaption}
\usepackage{bm}
\usepackage{bbm}
\usepackage{mathrsfs}
\usepackage{cleveref}
\usepackage{soul}
\usepackage{accents}
\usepackage{graphicx}
\usepackage{xcolor}
\usepackage{courier} 
\usepackage{listings} 
\usepackage{tabu} 
\usepackage{longtable}
\usepackage{changepage} 
\biboptions{sort&compress} 
\DeclareMathOperator\erf{erf} 

\journal{Corrosion Science}

\makeatletter
\def\@author#1{\g@addto@macro\elsauthors{\normalsize%
    \def\baselinestretch{1}%
    \upshape\authorsep#1\unskip\textsuperscript{%
      \ifx\@fnmark\@empty\else\unskip\sep\@fnmark\let\sep=,\fi
      \ifx\@corref\@empty\else\unskip\sep\@corref\let\sep=,\fi
      }%
    \def\authorsep{\unskip,\space}%
    \global\let\@fnmark\@empty
    \global\let\@corref\@empty  
    \global\let\sep\@empty}%
    \@eadauthor={#1}
}
\makeatother

\begin{document}

\begin{frontmatter}



\title{On the suitability of slow strain rate tensile testing for assessing hydrogen embrittlement susceptibility}


\author[IC]{Emilio Mart\'{\i}nez-Pa\~neda\corref{cor1}}
\ead{mail@empaneda.com}

\author[UVa]{Zachary D. Harris}

\author[Uniovi]{Sandra Fuentes-Alonso}

\author[UVa]{John R. Scully}

\author[UVa]{James T. Burns}

\address[IC]{Department of Civil and Environmental Engineering, Imperial College London, London SW7 2AZ, UK}

\address[UVa]{Department of Materials Science and Engineering, University of Virginia, Charlottesville, VA 22904, USA}

\address[Uniovi]{Department of Construction and Manufacturing Engineering, University of Oviedo, Gij\'{o}n 33203, Spain}

\cortext[cor1]{Corresponding author.}

\begin{abstract}
The onset of sub-critical crack growth during slow strain rate tensile testing (SSRT) is assessed through a combined experimental and modeling approach. A systematic comparison of the extent of intergranular fracture and expected hydrogen ingress suggests that hydrogen diffusion alone is insufficient to explain the intergranular fracture depths observed after SSRT experiments in a Ni-Cu superalloy. Simulations of these experiments using a new phase field formulation indicate that crack initiation occurs as low as 40\% of the time to failure. The implications of such sub-critical crack growth on the validity and interpretation of SSRT metrics are then explored.
\end{abstract}


\end{frontmatter}


\section{Introduction}
\label{Introduction}

Hydrogen (H)-induced, premature failure of structural metals undermines decades of metallurgical optimization, compromises structural integrity frameworks, and negatively affects industries spanning the aerospace, marine, transportation, and energy sectors \cite{Gangloff2003,Gangloff2008,Robertson2015,Lynch2019}. While the microscale mechanism(s) governing the H-embrittlement (HE) phenomenon continue to be debated \cite{Robertson2015,Lynch2019,Gerberich2012,AM2016,Harris2018}, it is universally recognized that HE must be explicitly accounted for when designing components that will be exposed to H-containing/producing environments and processes \cite{Gangloff2016}. Historically, the susceptibility of structural metals to environment-assisted cracking (EAC; such as HE) has been assessed via a diverse array of experimental protocols\footnote{For a complete listing of pertinent standards, see Ref. \cite{Raja2011}.} (e.g. NACE MR0175 \cite{NACEMR0175}, NACE TM0198 \cite{NACETM0198}, ASTM G30 \cite{ASTMG30}, ASTM G39 \cite{ASTMG39}, ASTM G129 \cite{ASTMG129}, ASTM E1681 \cite{ASTM1681}, ASTM F1624 \cite{ASTM1624}, and ISO 7539-7 \cite{ISO7539-7}), which are then utilized to inform a ``go/no-go'' decision regarding material selection. Specifically, if the evaluated material is found to exhibit susceptibility below an empirically defined 'threshold' parameter level for a given environment/loading combination \cite{Kim1979}, then the material is generally considered immune to EAC for the lifetime of the component.\\ 

The most commonly employed experiment for the assessment of EAC susceptibility is the slow strain rate test (SSRT), as demonstrated by two ASTM STP collections being dedicated to the method \cite{Ugiansky1979,Kane1993} and its documented use in $>40\%$ of EAC-related publications \cite{Henthorne2016}. The SSRT experimental protocol is codified in the ASTM G129 \cite{ASTMG129}, NACE TM0198 \cite{NACETM0198}, and ISO 7359-7 \cite{ISO7539-7} standards, and can be broadly described as a tensile experiment conducted using a slow, but constant extension rate (which typically yields an initial strain rate $<10^{-5}$ s$^{-1}$) while the specimen is exposed to the environment of interest \cite{Kim1979,Henthorne2016,McIntyre1988}. Numerous changes to this basic framework have been employed in the literature \cite{ASTMG129,Henthorne2016}. For example, specimens may be (1) exposed to the environment prior to testing (\emph{i.e.} pre-exposed or pre-charged) and/or during the experiment (\emph{in-situ}), (2) smooth or notched, (3) cyclically loaded prior to monotonic testing to generate a `fatigue pre-crack', and (4) modified to include specific features of interest (coatings \cite{Wright1995}, different microstructural zones induced by welding \cite{Theus1979,Klein1993}, tubular cross-sections \cite{Ahluwalia1993}, etc.). Metrics of interest depend on the specimen geometry \cite{Henthorne2016}, with reduction in area (i.e. ductility; also assessed \emph{via} strain-to-failure), fracture stress, and time-to-failure quantified for smooth specimens, while notch tensile strength and time-to-failure are typically evaluated for notched specimens.\\

The primary advantages of the SSRT approach are the simplicity and modularity of the experimental setup, the reduced cost relative to fracture mechanics-based approaches, and the reasonable test duration relative to static testing approaches \cite{Henthorne2016,Beavers1992}. Specifically, by virtue of the specimen being strained to failure, the SSRT approach avoids specifying a time for test completion, which is often a source of criticism for constant displacement and constant load-based testing strategies \cite{Beavers1992}. However, there are several complications that affect the efficacy of the SSRT method \cite{Kim1979,McIntyre1988,Beavers1992,Hibner1993}. First, the results of SSRT experiments have been reported to inadequately compare with in-service performance, with separate reports indicating that SSRT results are either overly conservative or not sufficiently conservative \cite{Kim1979,Beavers1992}. This limitation is explicitly incorporated into ASTM G129 \cite{ASTMG129}, which states that SSRT ``results are not intended to necessarily represent service performance'' and that the test method is only meant to act as a screening process to identify susceptible materials. These challenges in assessing susceptibility relative to in-service conditions are further complicated by the significant scatter in obtained SSRT metrics \cite{Hibner1993,Pollock1992}. Second, depending on the environment of interest, SSRT metrics have been shown to be strongly dependent on testing variables and specimen geometry. In particular, SSRT results have been found to be affected by the applied displacement rate \cite{Mclntyre1988,Lee2012,Margot-Marette1987}, the specimen surface finish \cite{Ahluwalia1993,Hibner1993,Hong2001}, and the specimen diameter \cite{Mclntyre1988}. For example, McIntyre noted that a slight change in strain rate from $4.0 \times 10^{-6}$ to $3.5 \times 10^{-6}$ s$^{-1}$ resulted in a 14\% decrease in time to failure for testing in a simulated sour gas environment (alloy not specified) \cite{Mclntyre1988}. Lastly, and perhaps most critically, if sub-critical crack growth (\emph{i.e.} crack growth (or the growth of multiple cracks) that does not immediately lead to failure) occurs during the SSRT experiment, then the applied remote stress (or nominal net-section stress if the sample is notched) is no longer representative of the operative mechanical driving force (which would instead be described by a \emph{stress intensity} \cite{Anderson2005,HenryHolroyd2019}).  Clear experimental evidence of such crack growth has been documented by Holroyd and coworkers in 5xxx-series Al alloys \cite{HenryHolroyd2019,Holroyd2017,Seifi2016}. This sub-critical crack growth will also obfuscate the physical meaning of other traditional SSRT metrics, such as time to failure and elongation. For example, the time to failure under such a scenario would be comprised of both the time for crack initiation and crack propagation \cite{Ahluwalia1993}. However, despite significant evidence of sub-critical cracking in the SSRT literature \cite{Ahluwalia1993,HenryHolroyd2019,Holroyd2017,Seifi2016,Haruna1994,Garud1990,Mathis2011,Sampath2018}, studies that systematically assess (1) the onset of sub-critical cracking during SSRT experiments and (2) the implications of such cracking on the interpretation of SSRT metrics are limited.\\

The objective of this study is to provide evidence that sub-critical crack growth can occur during typical SSRT experiments and then inform the possible implications of this crack growth on the interpretation of SSRT results. Recent efforts examining the H environment-assisted cracking (HEAC) behavior of a peak-aged Ni-Cu superalloy, Monel K-500, using both SSRT and linear elastic fracture mechanics (LEFM)-based approaches offers a well-characterized platform to assess such effects \cite{RinconTroconis2017,Harris2016}. The potential onset of sub-critical crack growth prior to failure during SSRT testing is evaluated based on systematic comparisons between expected H diffusion distances and the extent of intergranular cracking on the fracture surface of SSRT specimens. Based on this evidence, the time for crack initiation is estimated using an experimentally-calibrated phase field model for H-induced cracking. The results of these calculations are then utilized to discuss how the interpretation of SSRT metrics could be compromised by this sub-critical cracking phenomenon and to inform possible testing strategies to avoid this experimental complication.

\section{Experimental Methods}
\label{Sec:Experiments}

\subsection{Materials}

Monel K-500 is a precipitation-hardened, Ni-base alloy nominally comprised of a face-centered cubic (fcc) Ni-Cu solid solution ($\gamma$) matrix and a homogeneous distribution of highly coherent ($<0.1$ pct misfit strain), intermetallic $\gamma'$ (Ni3(Al,Ti)) precipitates \cite{Dey1986,Shoemaker2006,Dey1993,Davis2000}. The $\gamma'$ precipitates have an ordered Ll$_2$ structure composed of Ni atoms at the faces and Al (or Ti) atoms on the corners of the unit cell \cite{Dey1986,Davis2000}. Due to the low misfit and interfacial energy anisotropy between the $\gamma$ and $\gamma'$ phases, the precipitates form as spherical particles \cite{Dey1986}. Two types of carbides have been observed in Monel K-500: a heterogeneous distribution of MC-type carbides (typically TiC) in the $\gamma$ matrix and isolated M$_{23}$C$_6$-type carbides (where M: Cr, Mn, Fe, or Ni) on grain boundaries \cite{Dey1986}. Regarding the former, literature indicates that the TiC distribution does not appreciably change under typical aging conditions for Monel K-500 \cite{Gustafson2000} and that their contribution to the strength of Monel K-500 is minimal \cite{Dey1986}.\\

Four Monel K-500 material heats were evaluated in this study. One heat (termed Allvac) was supplied by Allegheny Technologies Incorporated and underwent the following aging protocol: direct aging at 593$^\circ$C (866 K) for 16 h, followed by a furnace cool at 14$^\circ$C/h to 482$^\circ$C (755 K) and then air cooling. A second heat (termed TR2) was harvested from an engineering component after exposure to a marine environment for 10-15 years; the exact heat treatment and exposure time/conditions are not known for this material heat. However, in order to enter service, the material heat was required to pass the QQ-N-286G federal procurement specification \cite{QQ-N-286G}, which mandates heat-treating to the near-peak aged condition. The final heats, supplied by the U.S. Naval Research Laboratory, were heat-treated \cite{Bayles2010} to obtain the lower and upper bound strengths possible under the QQ-N-286G federal specification \cite{QQ-N-286G}; denoted as NRL LS (low strength) and NRL HS (high strength), respectively. NRL LS and NRL HS were received in the form of 10.16-cm and 11.29-cm diameter barstock and then thermomechanically processed as follows \cite{Bayles2010}. NRL LS was hot rolled, continuously annealed at 982$^\circ$C (1255 K) followed by water quench, rotatory straightened, rough turned (6.35 mm removed from diameter), 3-point straightened, and then aged at 593$^\circ$C (866 K) for 2 h, followed by a furnace cool to 482$^\circ$C (755 K) at 55$^\circ$C/h and then air cool. The NRL HS was direct-aged at 593$^\circ$C (866 K) for 16 h, furnace cooled at 14$^\circ$C/h up to 538$^\circ$C (811 K), held for 1 h then furnace cooled at 14$^\circ$C/h to 482$^\circ$C (755 K), held for 1 h, and then air cooled. The bulk and trace composition for each material heat are given in Table \ref{Tab:Composition}.

\begin{table}[H]
\begin{adjustwidth}{-2cm}{}
\centering
\caption{Composition of evaluated Monel K-500 material heats.}
\label{Tab:Composition}
   {\tabulinesep=1.2mm
   \begin{tabu} {|c|c|c|c|c|c|c|c|c|c|c|c|c|c|c|}
       \hline
 & \textbf{Ni} & \textbf{Cu} & \textbf{Al} & \textbf{Fe} & \textbf{Mn} & \textbf{Si} & \textbf{Ti} & \textbf{C} & \textbf{S} & \textbf{P} & \textbf{Sn} & \textbf{Pb} & \textbf{Mg} & \textbf{Zr}\\ \hline
 \textbf{Heat} & \multicolumn{8}{c|}{\emph{{Weight percent (wt. \%)}}} & \multicolumn{6}{c|}{\emph{Weight parts per million (wppm)}}  \\ \hline
 Allvac & 66.12 & 28.57 & 2.89 & 0.80 & 0.81 & 0.08 & 0.45 & 0.17 & 1.6 & 92 & 2.4 & 2.1 & 39 & 370 \\ \hline
 TR2 & 64.66 & 30.15 & 2.73 & 0.69 & 0.73 & 0.09 & 0.45 & 0.20 & 11.0 & 71 & 6.9 & 2.5 & 130 & 330 \\  \hline
 NRL LS & 63.06 & 30.67 & 3.46 & 1.27 & 0.78 & 0.08 & 0.47 & 0.14 & 0.92 & 56 & 1.4 & 4.8 & 210 & 230 \\ \hline
 NRL HS & 63.44 & 30.74 & 3.20 & 0.91 & 0.85 & 0.10 & 0.57 & 0.14 & 17.0 & 40 & 2.2 & 3.5 & 40 & 650\\
\hline
   \end{tabu}}
 \end{adjustwidth}  
\end{table}

Uniaxial tension tests are conducted to characterize the elastic-plastic response of each material heat. The experimental results are then fitted to a Ramberg-Osgood power law hardening relationship \cite{Anderson2005}, as
\begin{equation}\label{eq:RambergOsgood}
    \varepsilon=\frac{\sigma}{E} + \alpha \frac{\sigma}{E} \left( \frac{\sigma}{\sigma_Y}\right)^{n}
\end{equation}

\noindent where $E$ is Young's modulus, $\sigma_Y$ is the yield stress, $n$ is the strain hardening exponent and $\alpha$ is the Ramberg-Osgood hardening parameter. The values measured are reported in Table \ref{Tab:MaterialProperties}, along with the average grain size for each material heat. 

\begin{table}[H]
\centering
\caption{Material properties.}
\label{Tab:MaterialProperties}
   {\tabulinesep=1.2mm
   \begin{tabu} {|c|c|c|c|c|c|c|}
       \hline
Heat & $E$ (GPa) & $\nu$ & $\sigma_Y$ (MPa) & $n$ & $\alpha$ & avg. grain size ($\mu$m) \\ \tabucline[1 pt]{1-7}
 Allvac   & 180 & 0.3 & 794.3 & 20 & 0.39 & 13.8  \\ \hline
 TR2 & 202 & 0.3 & 795 & 22 & 0.385 & 35.3 \\ \hline
 NRL LS & 198 & 0.3 & 715.7 & 21 & 0.5 & 22.5 \\ \hline 
 NRL HS & 191 & 0.3 & 910.1 & 18 & 0.405 & 11.2 \\
\hline
   \end{tabu}}
\end{table}

The diffusible H concentration, $C_{H,Diff}$, for each tested material heat was previously assessed using electrochemical extraction \cite{Ai2013,RinconTroconis2017}. Briefly, these electrochemical extraction measurements were completed on samples that were pre-charged for ten days at potentials ranging from -0.7 to -1.2 V$_{SCE}$ in 0.6 M NaCl with pH adjusted to 8.0 using NaOH. Extraction measurements were taken immediately after H charging \textit{via} H oxidation at -0.85 V vs. mercury-mercury sulfate electrode (MMSE) in deaerated borate buffer at pH 10 under ambient temperature conditions \cite{Ai2013}. For additional details of the experimental procedure for obtaining $C_{H,Diff}$, see Refs. \cite{Ai2013,RinconTroconis2017}. The values of $C_{H,Diff}$ measured for each material heat are given as a function of the applied potential $E_A$ in Table \ref{Tab:DiffusibleH}.

\begin{table}[H]
\centering
\caption{Diffusible hydrogen concentration $C_{H,Diff}$ (in wppm) for each Monel K-500 heat versus applied potential $E_A$.}
\label{Tab:DiffusibleH}
   {\tabulinesep=1.2mm
   \begin{tabu} {|c|c|c|c|}
       \hline
Heat & $E_A=-850$ mV & $E_A=-950$ mV & $E_A=-1100$ mV \\ \tabucline[1 pt]{1-4}
 Allvac & 1.9 & 4.1 & 7.5  \\ \hline
 TR2 & 3.7 & 18.6 & 26.2   \\ \hline
 NRL LS & 1.3 & 4.7 & 18.3  \\ \hline 
 NRL HS & 4.7  & 11.9  & 23.4  \\ 
\hline
   \end{tabu}}
\end{table}

\subsection{Slow strain-rate testing and characterization}

\begin{figure}[H]
\centering
\includegraphics[scale=0.4]{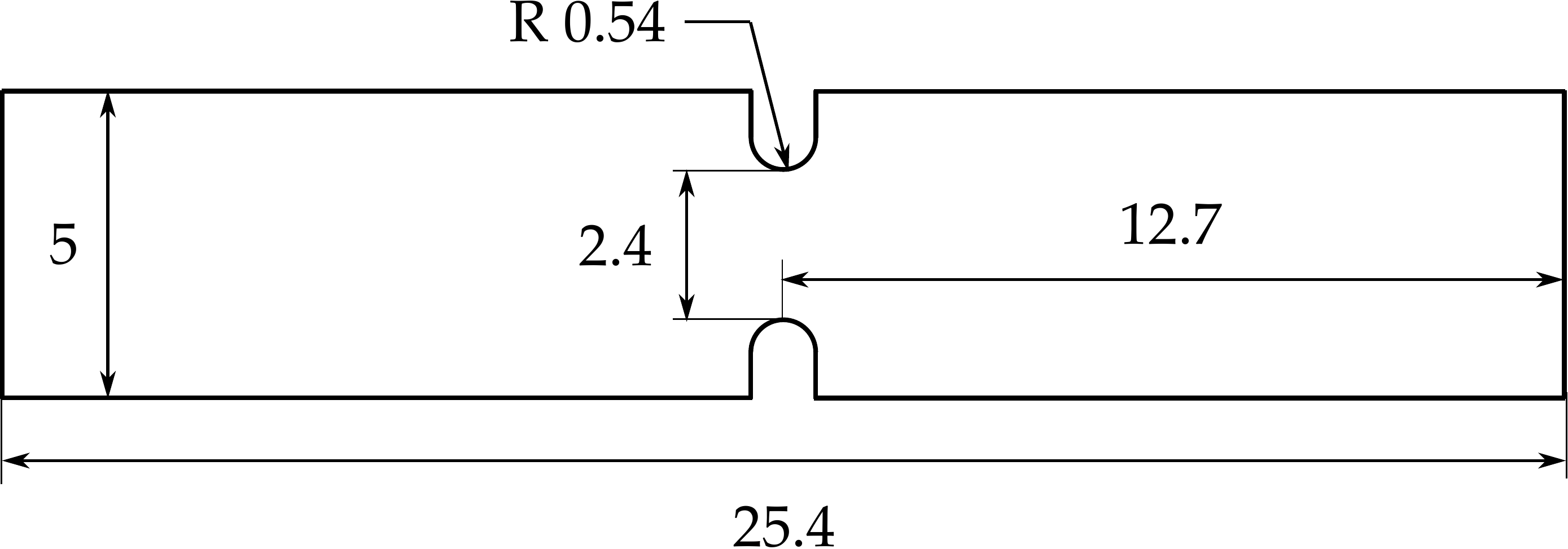}
\caption{Dimensions (in mm) of the specimens tested.}
\label{fig:Specimen}
\end{figure}

The SSRT experiments modeled in this study were completed using a Cortest, Inc servo-electric mechanical frame operated at an average cross-head speed of 6 $\times$ 10$^{-6}$ mm/s. Some subtle variations in the cross-head speed were noted across the test matrix (range was {4-8 $\times$} 10$^{-6}$ mm/s), but the cross-head speed within a given test was observed to be nominally constant. The cylindrical SSRT specimens were machined with a circumferential notch with dimensions shown in Figure \ref{fig:Specimen}.  SSRT testing was completed with the specimen fully immersed in deaerated 0.6 M NaCl (pH of 5.5-6.0) at 25$^\circ$C using a cylindrical Plexiglass cell. For each SSRT experiment, the specimen was baked at 450$^\circ$C for 48 h to remove any residual H and then placed within the testing cell. Then, prior to the onset of straining, each specimen was pre-charged with H at the testing potential for 48 h using a three-electrode setup, with a Pt mesh as the counter electrode, the specimen as the working electrode, and a saturated calomel electrode (SCE) as the reference electrode. It should be noted that this charging time was not sufficient to achieve H saturation across the specimen thickness, regardless of the applied potential. Each material heat was tested at three applied potentials (-0.95, -1.0, and -1.1 V$_{SCE}$) and the potential for a given experiment was also applied throughout the duration of the SSRT. It should be understood that the use of a constant applied potential results in a fixed H surface coverage, which is in equilibrium with the lattice H concentration. This lattice H concentration in turn determines the diffusible H concentration that participates in the H-induced fracture process \cite{Gangloff2014}. Comparison SSRT experiments were completed in laboratory air; these samples were also baked prior to straining for 48 h at 450$^\circ$C to ensure the removal of any residual H, but were not pre-charged with H. Additional details of the SSRT experiments are presented elsewhere \cite{RinconTroconis2017}.\\

Overview fractography was captured for each specimen using a FEI Quanta 650 FEG scanning electron microscope (SEM) after test completion to quantify the extent of intergranular cracking on the fracture surface. From these images, the 2D area of the fracture surface containing evidence of ductile failure was measured using the ImageJ software package and then utilized to calculate an effective ductile area radius. As the ductile region occurred in the center of the SSRT specimen, the intergranular crack penetration distance can then be approximated by subtracting this effective radius from the total radius of the specimen. 

\section{Numerical model}
\label{Sec:FEM}

Hydrogen transport towards the notch tip and subsequent cracking are investigated by means of a coupled deformation-diffusion-phase field fracture finite element framework. The implementation follows the recent work by Mart\'{\i}nez-Pa\~neda \textit{et al.} \cite{CMAME2018}, and extends it to account for the role of plasticity by means of $J_2$ flow theory. For simplicity, we choose to neglect the stress elevation associated with plastic strain gradients, which is notably less prominent in notches than in cracks \cite{IJP2016,TAFM2017}. However, gradient effects become relevant with crack advance and play an important role in understanding crack growth from existing sub-micron flaws \cite{JMPS2019}; the implications of adopting conventional plasticity theory will be subsequently discussed in the context of the results.\\

\subsection{Phase field fracture of embrittled elastic-plastic solids}
\label{Sec:PhaseField}

Consider an elastic-plastic solid occupying the domain $\Omega$ with a discontinuous surface $\Gamma$. The fracture resistance of the solid is characterized by a critical fracture energy $G_0$, which is is a macroscopic variable, akin to the fracture toughness, and can be employed to phenomenologically capture any damage mechanism. In the limiting case of an ideally brittle solid, $G_0$ corresponds to the Griffith critical energy release rate. As Monel K-500 exhibits predominantly intergranular fracture in the presence of H \cite{RinconTroconis2017,Harris2016,Harris2013,Gangloff2014,Burns2016a,Efird1985,Wolfe1988,Wolfe1990}, it is assumed that $G_0$ is dependent on the H coverage at the grain boundary $\theta$, which can be related to the bulk H concentration $C$ through the Langmuir-McLean isotherm:
\begin{equation}\label{Eq:Langmuir-McLean}
    \theta=\frac{C}{C+\exp \left( \frac{-\Delta g_b^0}{RT} \right)}
\end{equation}

\noindent Here, $C$ is given in units of impurity mol fraction, $R$ is the universal gas constant, $T$ is the temperature, and $\Delta g_b^0$ is the binding energy for the impurity at the site of interest. We follow Serebrinsky \textit{et al.} \cite{Serebrinsky2004} and assume $\Delta g_b^0=30$ kJ/mol for H trapped at a grain boundary. While it is anticipated that variations in grain boundary character will give rise to a distribution of trap binding energies, the use of 30 kJ/mol in the current study as an average value for H trapped at grain boundaries is justified based on experimental and computational studies \cite{Bhadeshia2016,Pressouy1981,Huang2017}. Assuming that H linearly degrades the fracture energy, the variation in $G_0$ with grain boundary H coverage can be defined as:
\begin{equation}\label{eq:G0X}
    G_0 \left( \theta \right) = \left( 1 - \chi \theta \right) \, G_0 \left( 0 \right)
\end{equation}

\noindent where $\chi$ is the H damage coefficient, which describes the potency of a unit coverage of H in degrading the fracture resistance of a material. This parameter may be calibrated with experimental data or connected with variables characterizing the underlying physical mechanism(s). For example, in the case of H-enhanced grain boundary decohesion, it can be inferred from atomistic calculations. The assumption of a linear degradation is supported by recent density function theory (DFT) analyses of H-induced decohesion in nickel, which observed a linear decrease in the surface energy with increasing H coverage \cite{Alvaro2015}.\\

Under the phase field formulation, the discrete crack is approximated \textit{via} the phase field $\phi$, which is an auxiliary parameter that resembles the concept of a scalar damage variable in continuum damage mechanics. The phase field order parameter varies from 0 to 1, with $\phi=0$ describing an intact material and $\phi=1$ indicating a completely damaged material. The size of the regularized crack surface is governed by the choice of a phase field model-inherent length scale, defined by $\ell$. As shown by $\Gamma$-convergence, a regularized crack density functional $\Gamma_\ell \left( \ell, \phi \right)$ can then be defined, which converges to the functional of the discrete crack as $\ell \to 0$ \cite{Bourdin2008}. As such, the fracture energy due to the creation of a crack can be approximated as a volume integral
\begin{equation}
\int_\Gamma G_0 \left( C\right) \, \text{d} \Gamma  \approx \int_\Omega G_0 \left( C \right) \Gamma_{\ell} \left( \ell, \phi \right) \, \text{d} \Omega = \int_\Omega G_0 \left( C \right) \left( \frac{1}{2\ell}\phi^2 + \frac{\ell}{2} |\nabla \phi|^2 \right) \text{d} V
\label{eq:surfaceenergy}
\end{equation}

\noindent This regularized description of the crack topology makes the problem suitable for numerical analysis and enables the capturing of complex crack topologies and trajectories, as well as fracture phenomena such as crack nucleation, branching, and coalescence \cite{Wu2020}.\\ 

The total potential energy of a cracked body can be obtained from the sum of the surface energy associated with the formation of a crack $\Psi^s \left( \phi, C \right)$ and the bulk energy $\Psi^b \left( \bm{u}, \phi \right)$, where $\bm{u}$ is the displacement field:
\begin{equation}\label{eq:Psi2}
    \Psi = \Psi^b \left( \bm{u}, \phi \right) + \Psi^s \left( \phi, C \right)= \int_\Omega \left[  (1- \phi)^2 \, \psi \left( \bm{u} \right) + G_0 \left( C \right) \Gamma_\ell \left( \ell, \phi \right)  \right] \text{d} V
\end{equation}

\noindent Here, the bulk energy is given by the strain energy density $\psi \left( \bm{u} \right)$ of the elastic-plastic solid and a term accounting for the degradation of the stored energy with evolving damage. The strain energy density is additively decomposed into its elastic $\psi^e$ and plastic $\psi^p$ parts, such that $\psi =  \psi^e +\psi^p$. In previous studies, see Ref. \cite{CMAME2018} and the works by Duda and co-workers \cite{Duda2015,Duda2018}, fracture was assumed to be driven purely by the elastic strain energy density. However, here we follow Miehe \textit{et al.} \cite{Miehe2016b} and consider, for the first time in the context of hydrogen embrittlement, the contributions from both the elastic and plastic strain energy densities. In addition, we choose not to define an explicit relation between the plastic yield condition and the damage variable. No plastic-damage coupling is typically defined in similar fracture models, such as cohesive zone approaches (see, for example, Refs. \cite{Yu2016a,JAM2018}).\\
%

The strain tensor $\bm{\varepsilon}$ is computed from the displacement field in the usual manner $\bm{\varepsilon}= \text{sym} \nabla \bm{u}$ and additively decomposes into an elastic part $\bm{\varepsilon}^e$ and a plastic part $\bm{\varepsilon}^p$. The Cauchy stress tensor is defined as $\bm{\sigma}=\partial_\varepsilon \psi$. Taking the first variation of the total potential energy of the solid (Eq. (\ref{eq:Psi2})) with respect to $\bm{\varepsilon}$ and $\phi$ renders the weak form of the deformation-phase field fracture problem. Thus, in the absence of body forces and external tractions,
\begin{equation}\label{Eq:weak}
  \int_{\Omega} \left\{ \left( 1 - \phi \right)^2  \bm{\sigma} : \delta \bm{\varepsilon}  -2(1-\phi)\delta \phi \, \psi +
        G_0 \left( C \right) \left( \dfrac{\phi}{\ell} \delta \phi
        + \ell\nabla \phi \cdot \nabla \delta \phi \right) \right\}  \, \mathrm{d}V = 0
\end{equation}

Upon making use of Gauss' divergence theorem, the following coupled field equations are obtained for any arbitrary value of the kinematic variables $\delta \bm{u}$ and $\delta \phi$,
\begin{align}\label{eqn:strongForm}
(1-\phi)^2 \, \, \nabla \cdot \boldsymbol{\sigma}  &= \boldsymbol{0}   \hspace{3mm} \rm{in}  \hspace{3mm} \Omega \nonumber \\ 
G_{c} \left( C \right)  \left( \dfrac{\phi}{\ell}  - \ell \Delta \phi \right) - 2(1-\phi) \, \psi  &= 0 \hspace{3mm} \rm{in}  \hspace{3mm} \Omega
\end{align}

It is important to note that the phase field length scale $\ell$ is the material parameter that governs the critical stress at which damage initiates, which can be shown as follows. Consider the homogeneous solution to a one dimensional linear elastic solid with Young's modulus $E$ subjected to a strain $\varepsilon$. The strain energy reads $\psi=E \varepsilon^2 / 2$ and the homogeneous phase field can be readily obtained from Eq. (\ref{eqn:strongForm})b as
\begin{equation}
    \phi = \frac{E \varepsilon^2 \ell}{G_0 \left( C \right) + E \varepsilon^2 \ell}
\end{equation}

\noindent Thus, the effective stress $\bar{\sigma}=\left( 1 - \phi \right)^2 \sigma$ reaches a maximum at
\begin{equation}\label{eq:effectiveS}
    \bar{\sigma}_c = \left( \frac{27E G_0 \left( C \right)}{256 \ell} \right)^{1/2}
\end{equation}

As such, for a finite $\ell$, the phase field model resembles cohesive zone approaches, with the phase field length scale governing the size of the fracture process zone and the critical stress.

\subsection{Hydrogen transport}
\label{Sec:Htransport}

Mass conservation requirements relate the rate of change of the H concentration $C$ with the H flux $\bm{J}$ through the external surface as follows
\begin{equation}
    \int_{\Omega} \frac{dC}{dt} \, \mathrm{d}V + \int_{\partial \Omega} \bm{J} \cdot \bm{n} \, \mathrm{d}S = 0
\end{equation}

The strong form of the balance equation can be readily obtained by making use of the divergence theorem and noting that the expression must hold for any arbitrary volume,
\begin{equation}\label{Eq:StrongC}
    \frac{dC}{dt} + \nabla \cdot \bm{J} = 0
\end{equation}

For an arbitrary, suitably continuous, scalar field, $\delta C$, the variational statement Eq. (\ref{Eq:StrongC}) can be written as:
\begin{equation}\label{eq:Hweak0}
    \int_\Omega \delta C \left( \frac{dC}{dt} + \nabla \cdot \bm{J} \right) \, \mathrm{d}V  = 0
\end{equation}

Rearranging Eq. (\ref{eq:Hweak0}), and making use of the divergence theorem, the weak form of the balance equation can be obtained:
\begin{equation}\label{eq:Hweak1}
   \int_\Omega \left[ \delta C \left( \frac{dC}{dt} \right) - \bm{J} \cdot \nabla \delta C   \right] \, \mathrm{d}V + \int_{\partial \Omega_q} \delta C q \, \mathrm{d}S  = 0 
\end{equation}

\noindent where $q=\bm{J} \cdot \bm{n}$ is the concentration flux exiting the body across $\partial \Omega_q$. The diffusion is driven by the gradient of the chemical potential $\nabla \mu$. Accordingly, the mass flux follows a linear Onsager relationship,
\begin{equation}\label{Eq:Flux}
    \bm{J}= - \frac{D C}{R T} \nabla \mu
\end{equation}

\noindent where $D$ is the diffusion coefficient (under typical H-charging conditions, this would be the trap-modified effective H diffusivity $D_{eff}$). The chemical potential of H in lattice sites is given by:
\begin{equation}\label{Eq:ChePotential}
    \mu=\mu^0 + RT \ln \frac{\theta_L}{1-\theta_L} - \bar{V}_H \sigma_H
\end{equation}

\noindent where $\mu^0$ is the chemical potential in the standard case, $\theta_L$ is the occupancy of lattice sites, $\bar{V}_H$ is the partial molar volume of H in solid solution, and $\sigma_H$ is the hydrostatic stress. By substituting Eq. (\ref{Eq:ChePotential}) into Eq. (\ref{Eq:Flux}), and considering the relation between $\theta_L$ and the site density $N$, $\theta_L=C/N$, the flux can be described as follows:
\begin{equation}\label{eq:JfluxExtended}
   \bm{J}=- \frac{DC}{\left(1 - \theta_L \right)} \left(\frac{\nabla C}{C} - \frac{\nabla N}{N} \right) + \frac{D}{RT} C \bar{V}_H \nabla \sigma_H
\end{equation}

Now, substituting Eq. (\ref{eq:JfluxExtended}) into Eq. (\ref{eq:Hweak1}), and assuming low occupancy ($\theta_L \ll 1$) and a constant interstitial site concentration ($\nabla N=0$), the governing H transport equation becomes:
\begin{equation}\label{Eq:WeakC}
    \int_\Omega \left[ \delta C \left( \frac{1}{D} \frac{dC}{dt} \right) + \nabla \delta C  \nabla C  - \nabla \delta C \left( \frac{\bar{V}_H C}{RT}  \nabla \sigma_H \right) \right] \, \mathrm{d}V = - \frac{1}{D} \int_{\partial \Omega_q} \delta C q \, \mathrm{d}S
\end{equation}

Details of the coupling and the finite element implementation of the deformation-diffusion-fracture problem are given in \ref{App:FE}. Additionally, it is critical to accurately define the environmental conditions in the presence of a propagating crack. In the current study, a penalty approach is employed to capture how the electrochemical solution promptly occupies the space created with crack advance; details regarding implementation are provided in \ref{App:MovingBCs}.

\section{Results}
\label{Sec:Results}

\subsection{SSRT results and assessment of intergranular cracking depth as a function of applied potential and material heat}

\begin{figure}[H]
\makebox[\linewidth][c]{%
        \begin{subfigure}[b]{0.55\textwidth}
                \centering
                \includegraphics[scale=0.6]{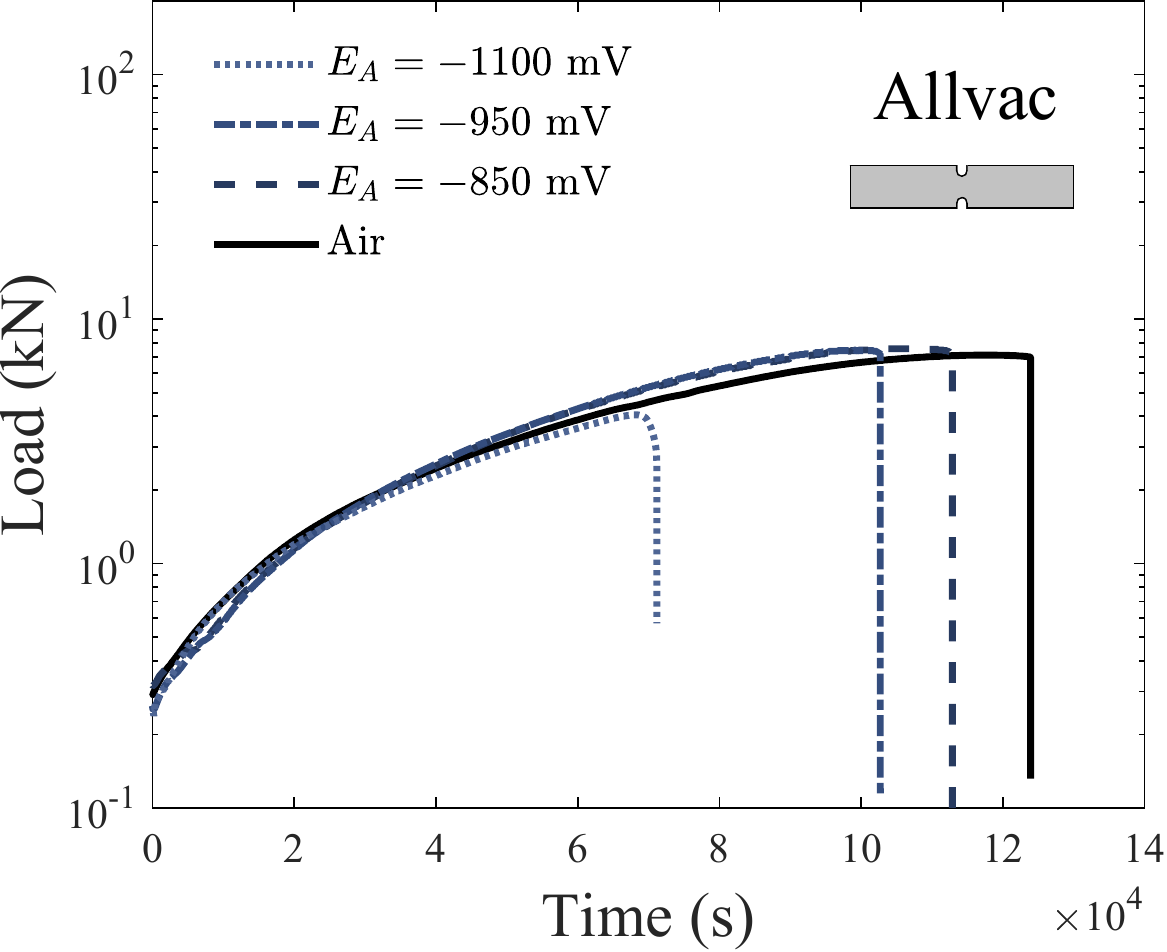}
                \caption{}
                \label{fig:AllvacFvsTime}
        \end{subfigure}
        \begin{subfigure}[b]{0.55\textwidth}
                \raggedleft
                \includegraphics[scale=0.6]{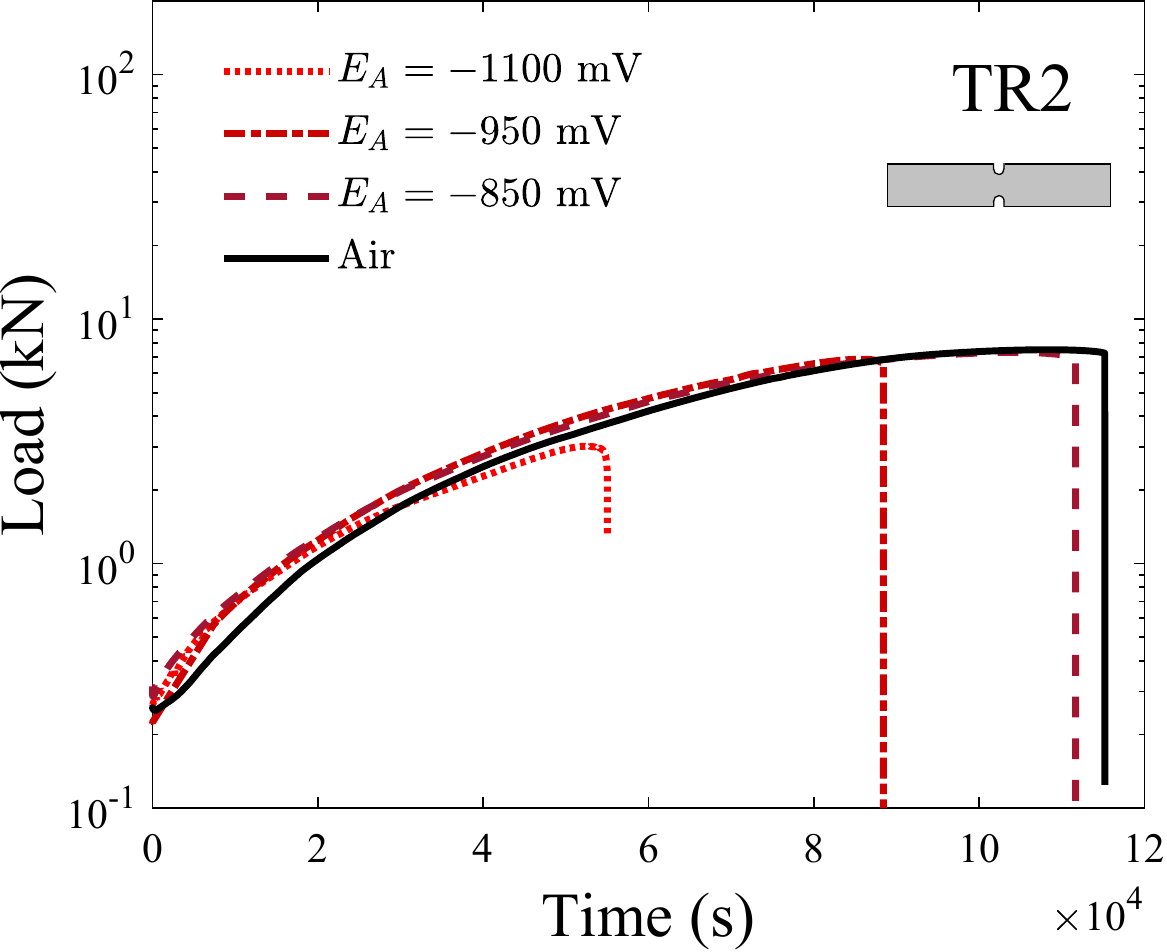}
                \caption{}
                \label{fig:TR2FvsTime}
        \end{subfigure}}

\makebox[\linewidth][c]{%
        \begin{subfigure}[b]{0.55\textwidth}
                \centering
                \includegraphics[scale=0.6]{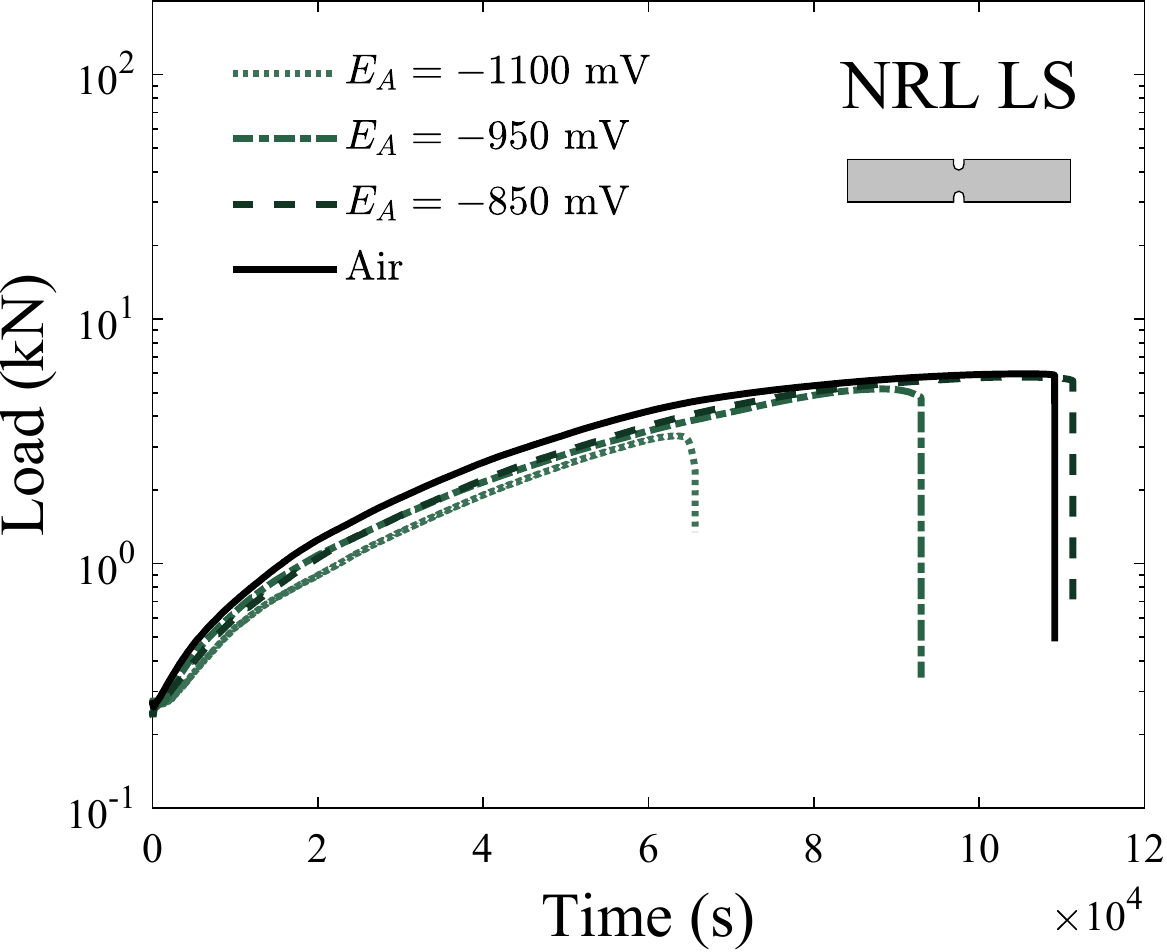}
                \caption{}
                \label{fig:NRLLSvsTime}
        \end{subfigure}
        \begin{subfigure}[b]{0.55\textwidth}
                \raggedleft
                \includegraphics[scale=0.6]{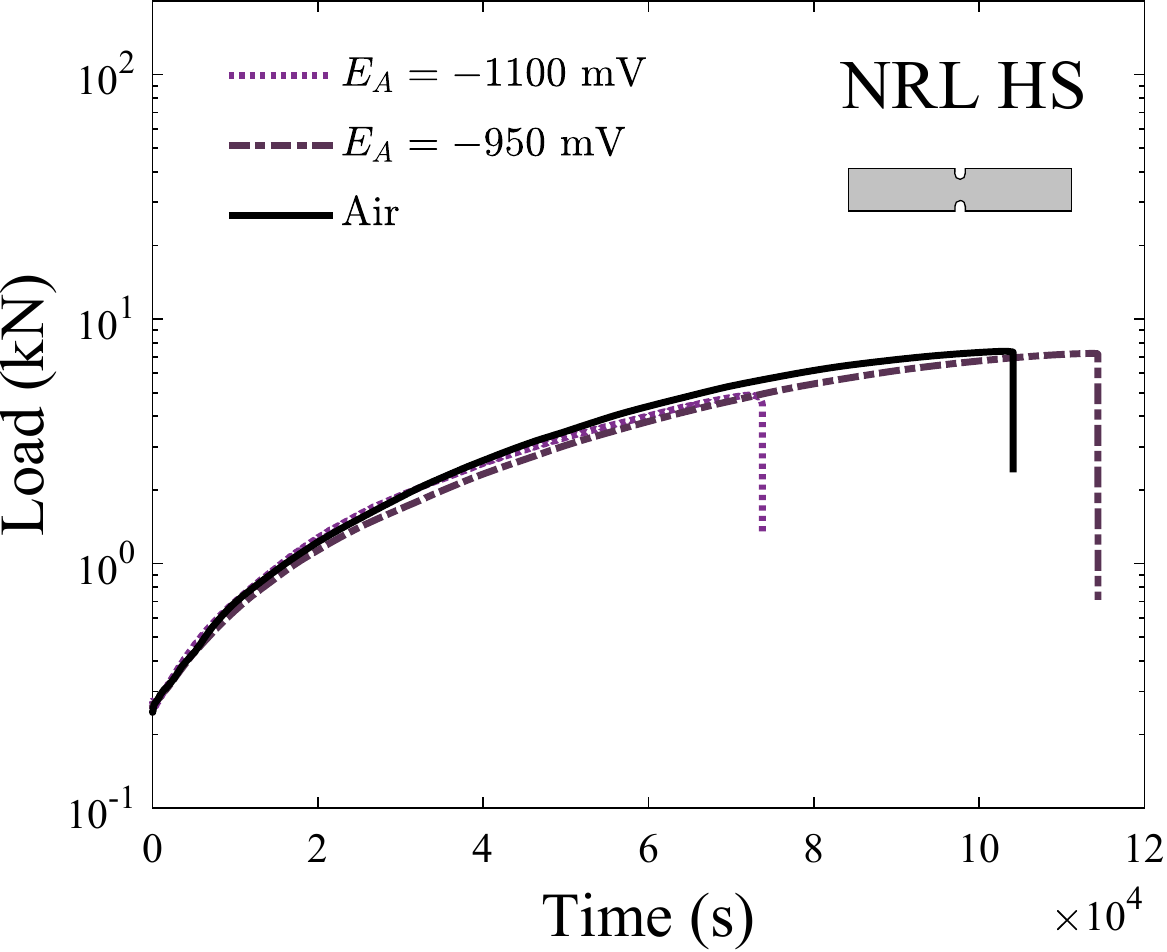}
                \caption{}
                \label{fig:NRLHSvsTime}
        \end{subfigure}
        }       
        \caption{Experimental load versus time results as a function of applied potential for the four evaluated material heats of Monel K-500: (a) Allvac, (b) TR2, (c) NRL LS, and (d) NRL HS.}\label{fig:ExptResults}
\end{figure}

The force-time data obtained from each SSRT experiment as a function of applied potential for the four tested Monel K-500 material heats are shown in Figure \ref{fig:ExptResults}. Consistent with prior fracture mechanics-based testing of HEAC susceptibility in Monel K-500 \cite{Harris2018,Harris2016,Gangloff2014,Burns2016a}, the time to failure was generally found to decrease across all tested material heats with increasingly negative applied potential; a similar trend was also noted for the applied load. As reported in prior work \cite{RinconTroconis2017}, all specimens exhibited intergranular fracture during testing at -1100 mV$_{SCE}$, though the center of the specimens failed via ductile microvoid coalesence. For potentials more positive than -1100 mV$_{SCE}$, the fracture morphology became increasingly different amongst the tested material heats. Specifically, neither Allvac or NRL HS contained any evidence of intergranular fracture at -950 or -850 mV$_{SCE}$, TR2 had intergranular fracture at -950 mV$_{SCE}$, but not -850 mV$_{SCE}$, and NRL LS exhibited intergranular fracture at both -950 and -850 mV$_{SCE}$. Such heat-to-heat variations in apparent HEAC susceptibility as a function of applied potential are consistent with prior fracture mechanics-based evaluations of Monel K-500 \cite{Harris2016}.

\begin{figure}[H]
\makebox[\linewidth][c]{%
        \begin{subfigure}[b]{0.65\textwidth}
                \centering
                \includegraphics[scale=0.32]{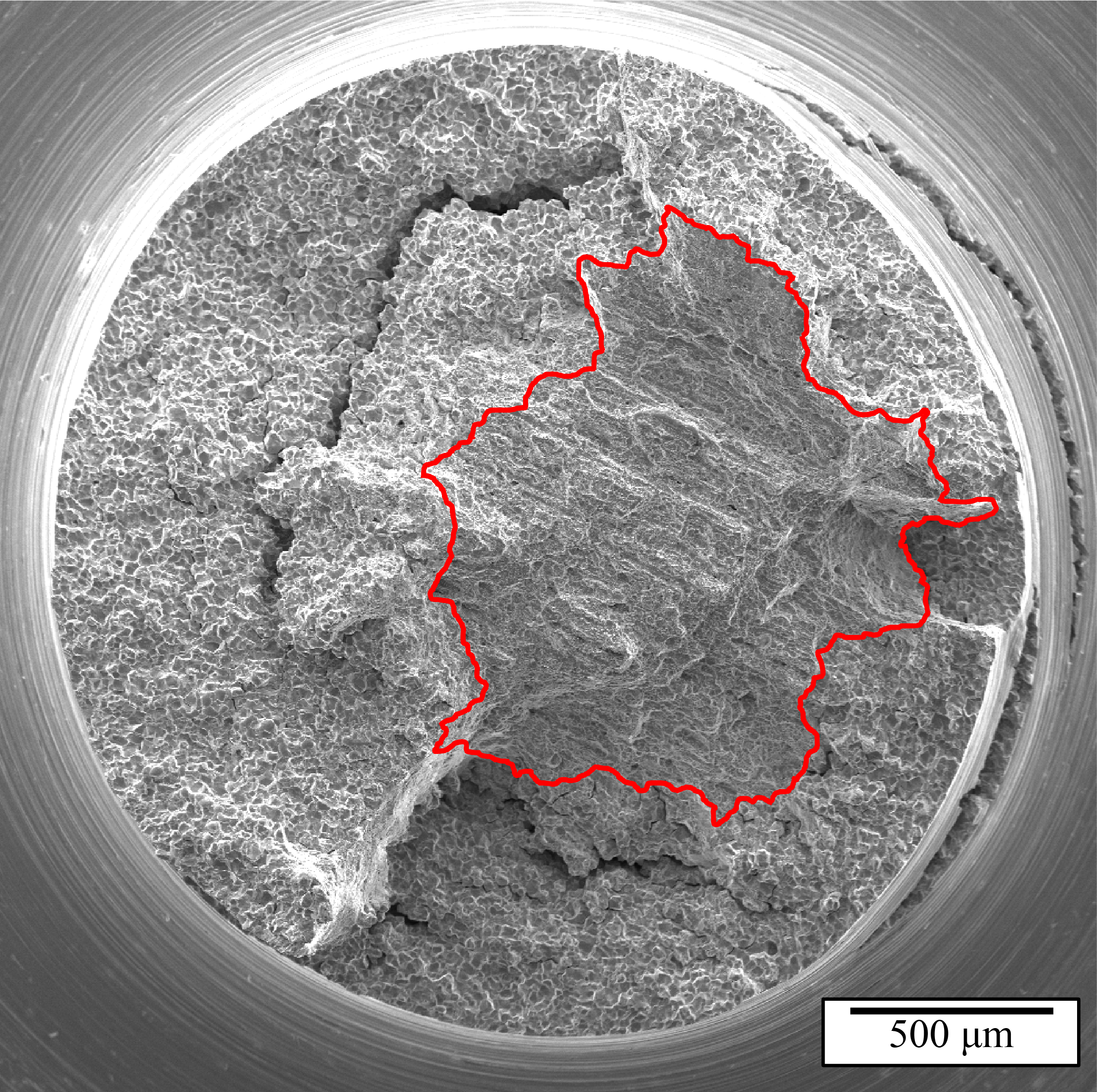}
                \caption{}
                \label{fig:Figure3a}
        \end{subfigure}
        \begin{subfigure}[b]{0.65\textwidth}
                \centering
                \includegraphics[scale=0.85]{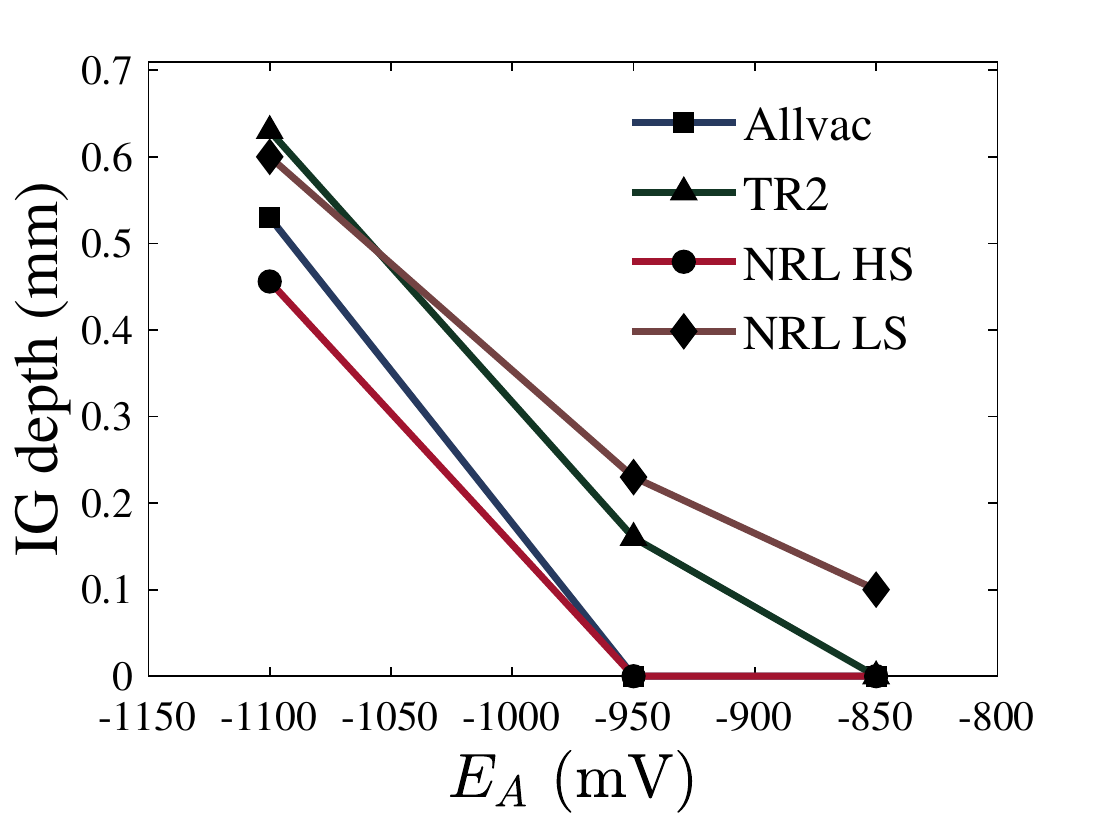}
                \caption{}
                \label{fig:Figure3b}
        \end{subfigure}}

\makebox[\linewidth][c]{%
        \begin{subfigure}[b]{0.65\textwidth}
                \centering
                \includegraphics[scale=0.47]{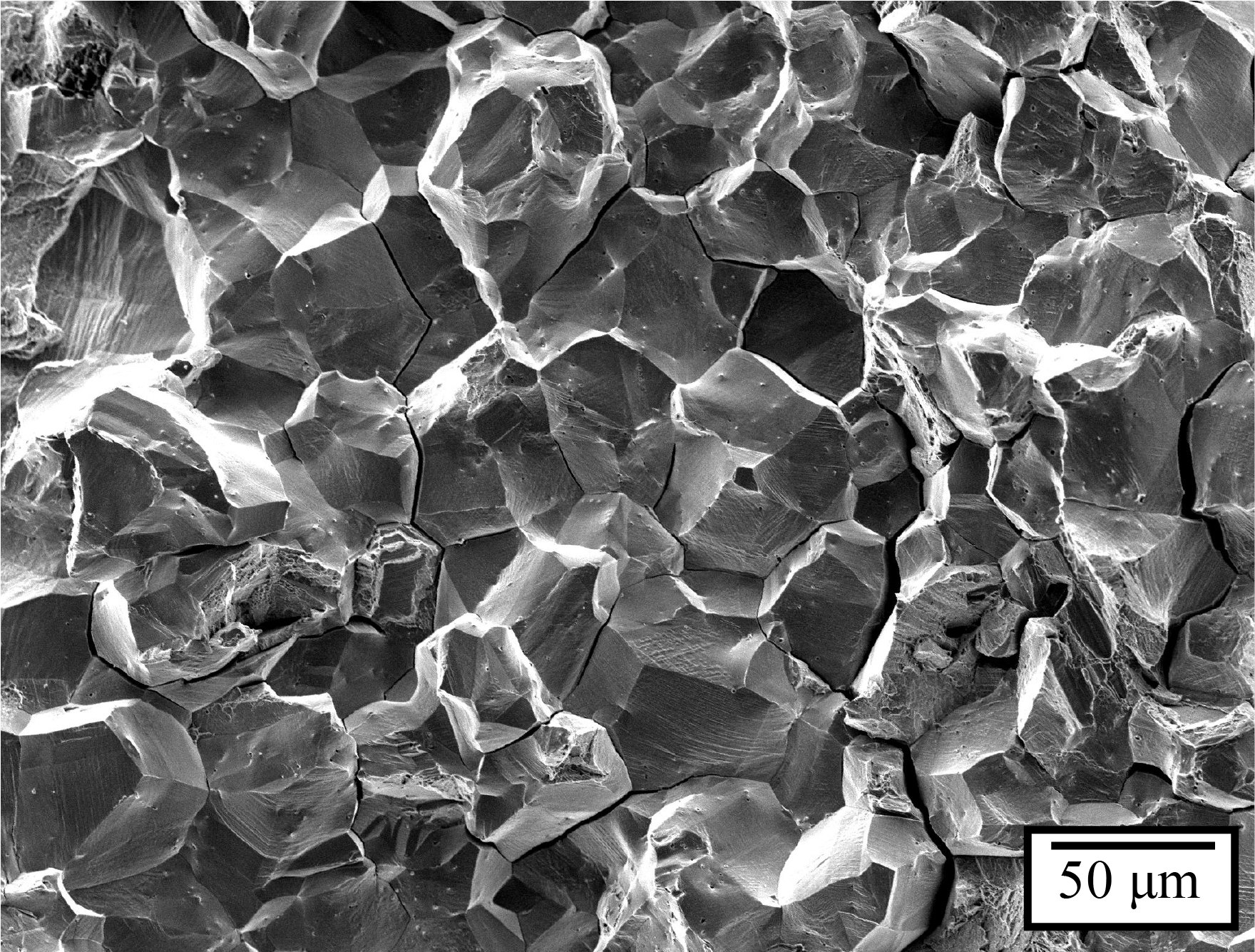}
                \caption{}
                \label{fig:Figure3c}
        \end{subfigure}
        \begin{subfigure}[b]{0.65\textwidth}
                \centering
                \includegraphics[scale=0.55]{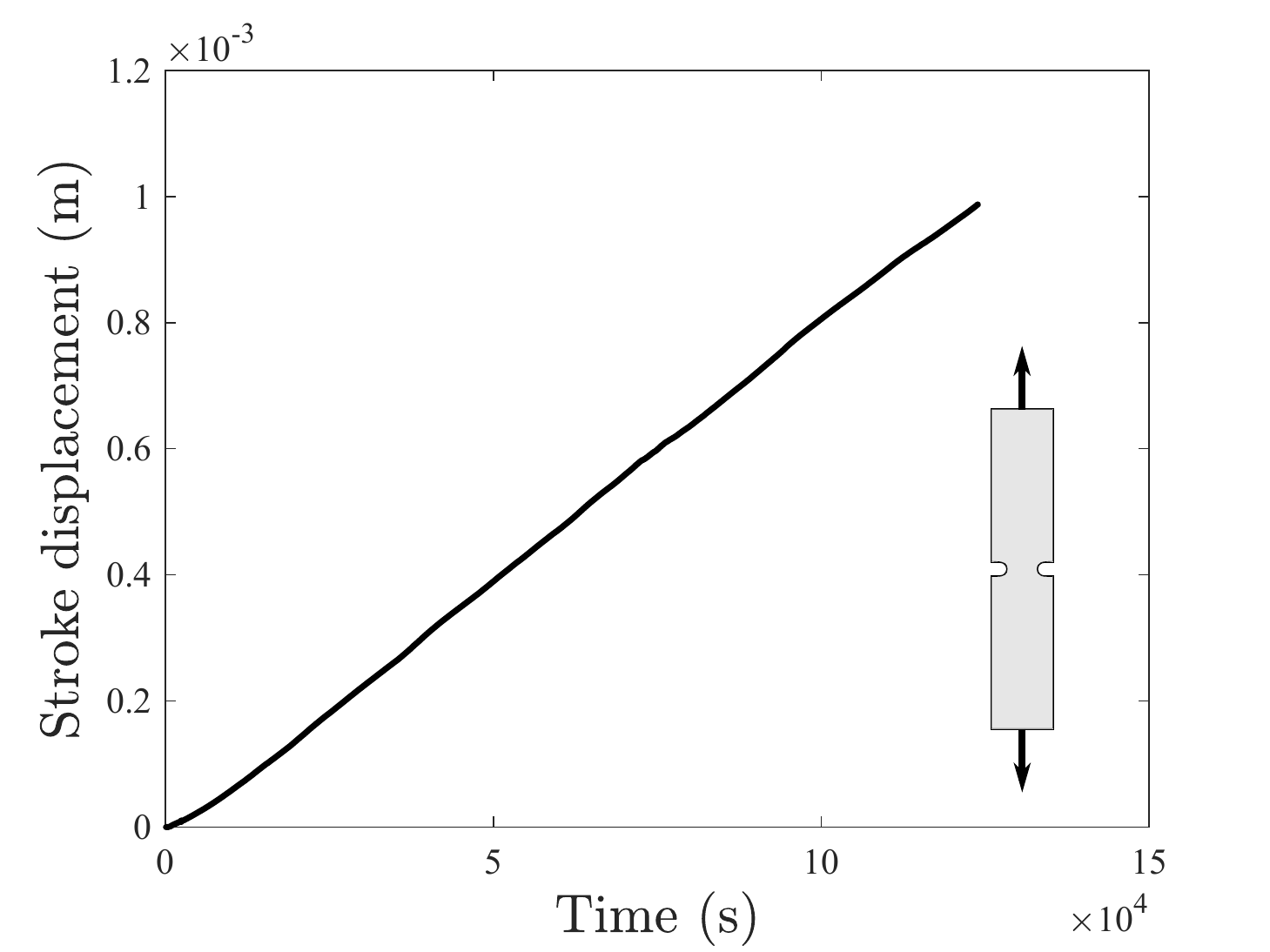}
                \caption{}
                \label{fig:Figure3d}
        \end{subfigure}
        }       
        \caption{(a) Overview micrograph of the fracture surface obtained for NRL LS tested at -1100 mV$_{SCE}$ with the region exhibiting ductile fracture outlined by the red dashed line. From these images, the effective depth of intergranular cracking (b) was determined for each of the material heats as a function of applied potential. A representative, higher magnification micrograph from TR2 tested at -1100 mV$_{SCE}$ of the observed intergranular fracture morphology in all tested Monel K-500 heats is shown in (c), while a representative stroke displacement versus time trace (d) taken from Allvac tested in air demonstrates the constancy of the cross-head speed during SSRT testing.}\label{fig:IGdepth}
\end{figure}

To assess the extent of intergranular fracture penetration as a function of applied potential and material heat, overview fractographs were collected for each tested specimen. These images were then utilized to determine the effective radius of the central region where ductile fracture was observed, which was then subtracted from the specimen radius to identify the effective depth of intergranular fracture. An example fractograph, from NRL LS tested at -1100 mV$_{SCE}$, illustrating this procedure (with the ductile core region delineated by the red line) is shown in Figure \ref{fig:Figure3a}, while the calculated depth of intergranular cracking as a function of material heat and applied potential are shown in Figure \ref{fig:Figure3b}. The depth of intergranular fracture was found to increase with increasingly negative applied potentials, consistent with the increased HEAC susceptibility previously measured in fracture mechanics-based experiments under such conditions \cite{Harris2016, Gangloff2014, Burns2016a}. Interestingly, this depth of integranular penetration was found to appreciably vary across the tested material heats, with NRL LS and TR2 consistently exhibiting increased integranular fracture depths relative to NRL HS and Allvac. A representative, high magnification micrograph of the observed intergranular fracture morphology, taken from the TR2 specimen tested at -1100 mV$_{SCE}$ is shown in Figure \ref{fig:Figure3c}. Lastly, given the sensitivity of SSRT metrics to the applied strain rate \cite{McIntyre1988}, it is important to establish that the cross-head speed does not appreciably vary during SSRT experiments. As demonstrated by the observed stroke displacement as function of testing time for the Allvac specimen tested in laboratory air, shown in Figure \ref{fig:Figure3d}, the cross-head speed throughout a given experiment was found to be nominally constant.

\subsection{Comparison between integranular cracking depth and hydrogen diffusion distance}

Examination of Figure \ref{fig:IGdepth}b shows that an intergranular fracture morphology was observed at depths ranging from approximately 0.45 to 0.65 mm for the four Monel K-500 material heats tested in 0.6 M NaCl at an applied potential of -1100 mV$_{SCE}$. Such differences in the depth of intergranular fracture across the four material heats are consistent with the heat-to-heat variations in HEAC susceptibility observed in fracture mechanics-based testing of Monel K-500 \cite{Harris2016}. These prior efforts also established that all evaluated heats of Monel K-500 failed via microvoid coalescence during inert environment testing, suggesting that H must be present for intergranular fracture to occur \cite{Harris2016,Gangloff2014,Burns2016a}. However, as addressed in detail below, non-negligible differences are found between the measured extent of intergranular cracking and the expected H diffusion distance.\\

The expected extent of H penetration into the specimen is assessed \emph{via} several avenues. First, assuming one-dimensional diffusion, the extent of H diffusion into the SSRT specimen can be estimated from the material diffusion coefficient by the following equation: 
\begin{equation}\label{eq:Diffdistance}
    d = 2 \sqrt{D \, t_f}
\end{equation}

\noindent where $d$ is the diffusion distance\footnote{It should be noted that the diffusion distance predicted by Eq. (\ref{eq:Diffdistance}) does not calculate the full extent of H diffusion, but instead represents the distance from the surface where the concentration ratio $C/C_b =1-\erf (1) \approx 0.157$ will be located after $t_f$ \cite{Shewmon2016}. However, the quantity of $2\sqrt{D t_f}$ is commonly considered a characteristic quantity for diffusion distance and is therefore utilized as a first-order approximation herein.} and $t_f$ is the time over which H was allowed to diffuse. Based on the work of Scully and co-workers \cite{RinconTroconis2017,Ai2013}, the H diffusivity for the NRL HS material heat was measured as $D=1.3 \times 10^{-10}$ cm$^2$/s. Using this diffusivity and $t_f = 245,960$ s (based on a 48-h pre-charge and 20.3-h test duration), the expected diffusion distance of H is 0.113 mm, which is $\sim 4 \times$ smaller than the experimentally observed intergranular crack depth for NRL HS tested at -1100 mV$_{SCE}$ (0.45 mm).\\ 

\begin{figure}[H]
\makebox[\linewidth][c]{%
        \begin{subfigure}[b]{0.55\textwidth}
                \centering
                \includegraphics[scale=0.7]{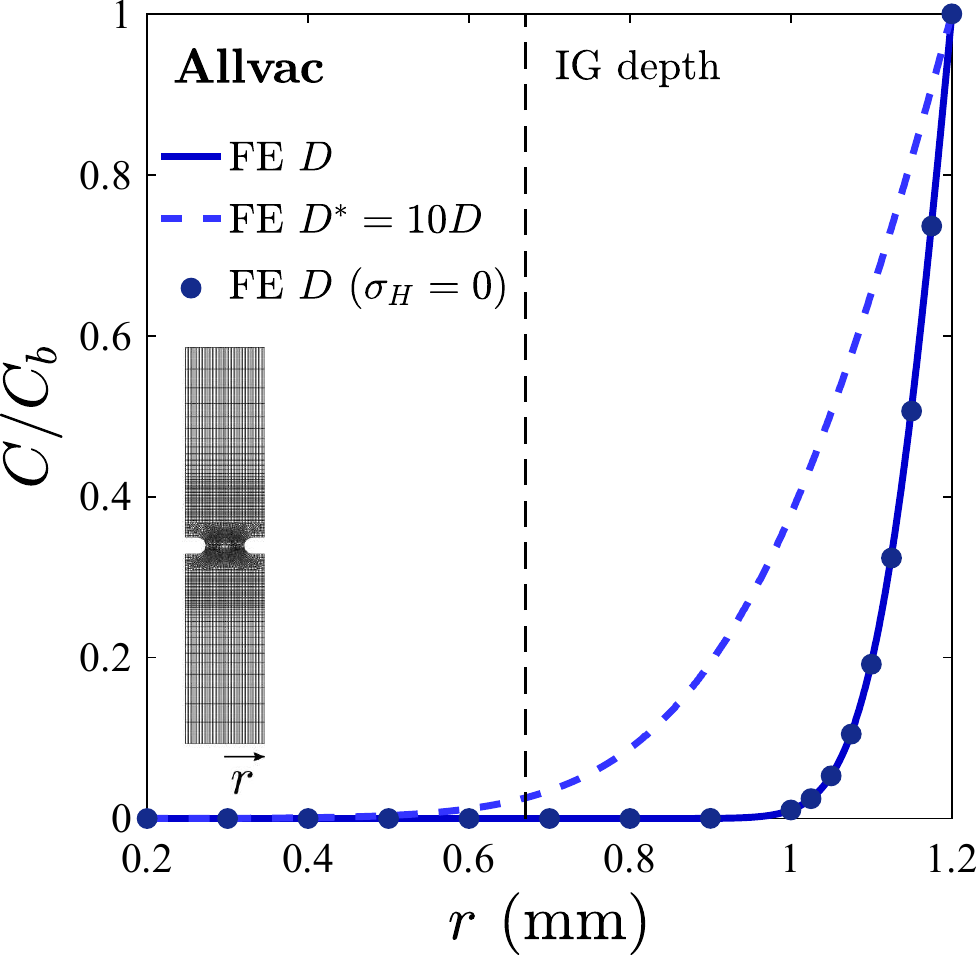}
                \caption{}
                \label{fig:AllvacDiff}
        \end{subfigure}
        \begin{subfigure}[b]{0.55\textwidth}
                \raggedleft
                \includegraphics[scale=0.7]{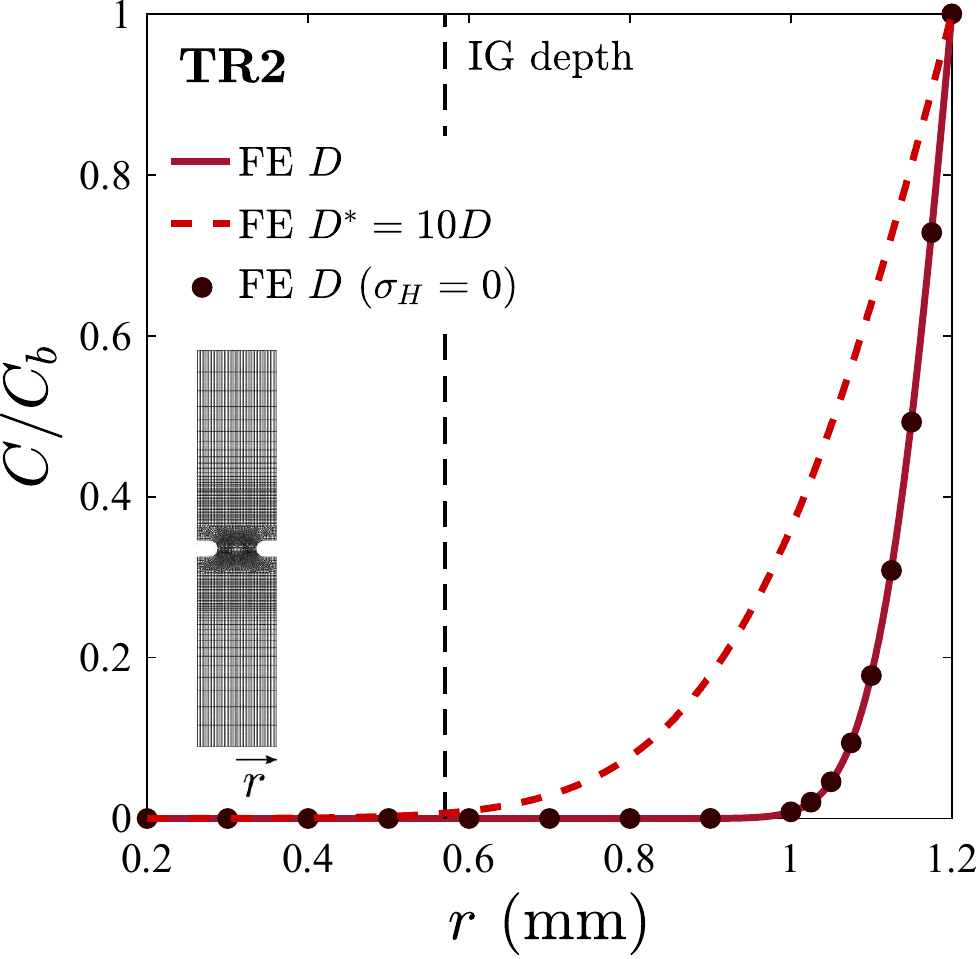}
                \caption{}
                \label{fig:TR2Diff}
        \end{subfigure}}

\makebox[\linewidth][c]{%
        \begin{subfigure}[b]{0.55\textwidth}
                \centering
                \includegraphics[scale=0.7]{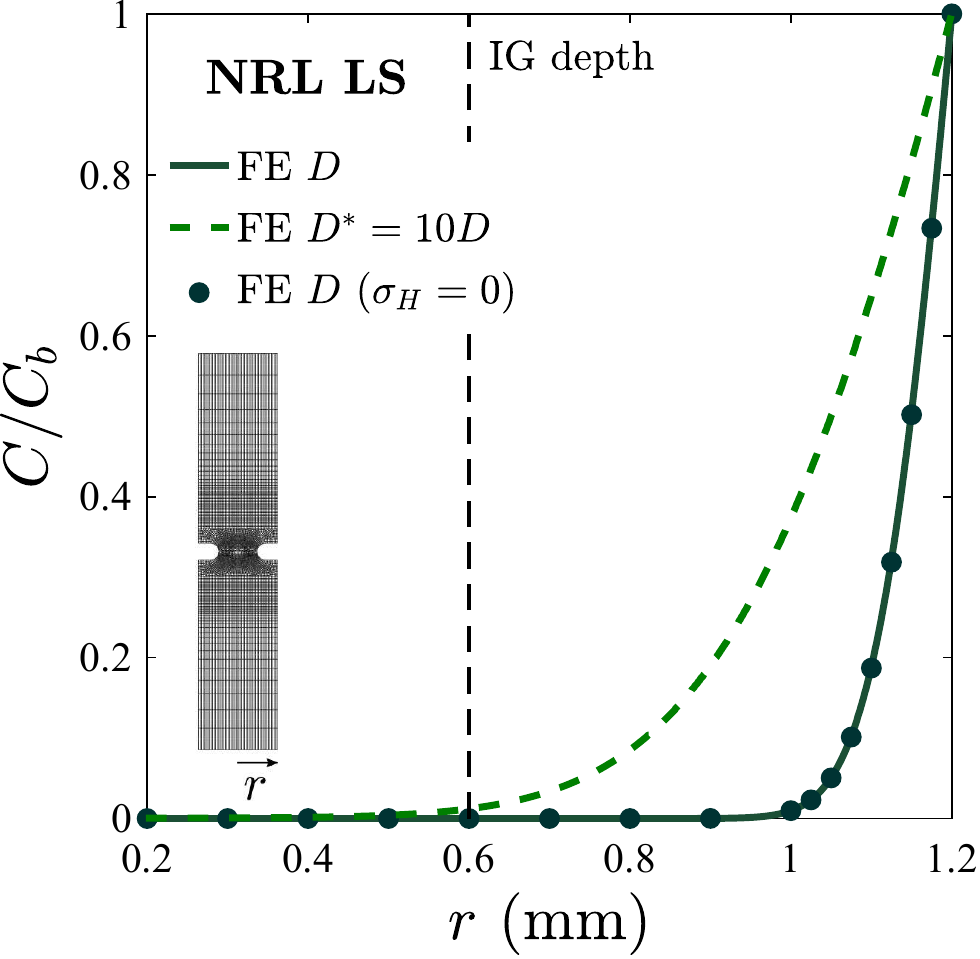}
                \caption{}
                \label{fig:NRLLSDiff}
        \end{subfigure}
        \begin{subfigure}[b]{0.55\textwidth}
                \raggedleft
                \includegraphics[scale=0.7]{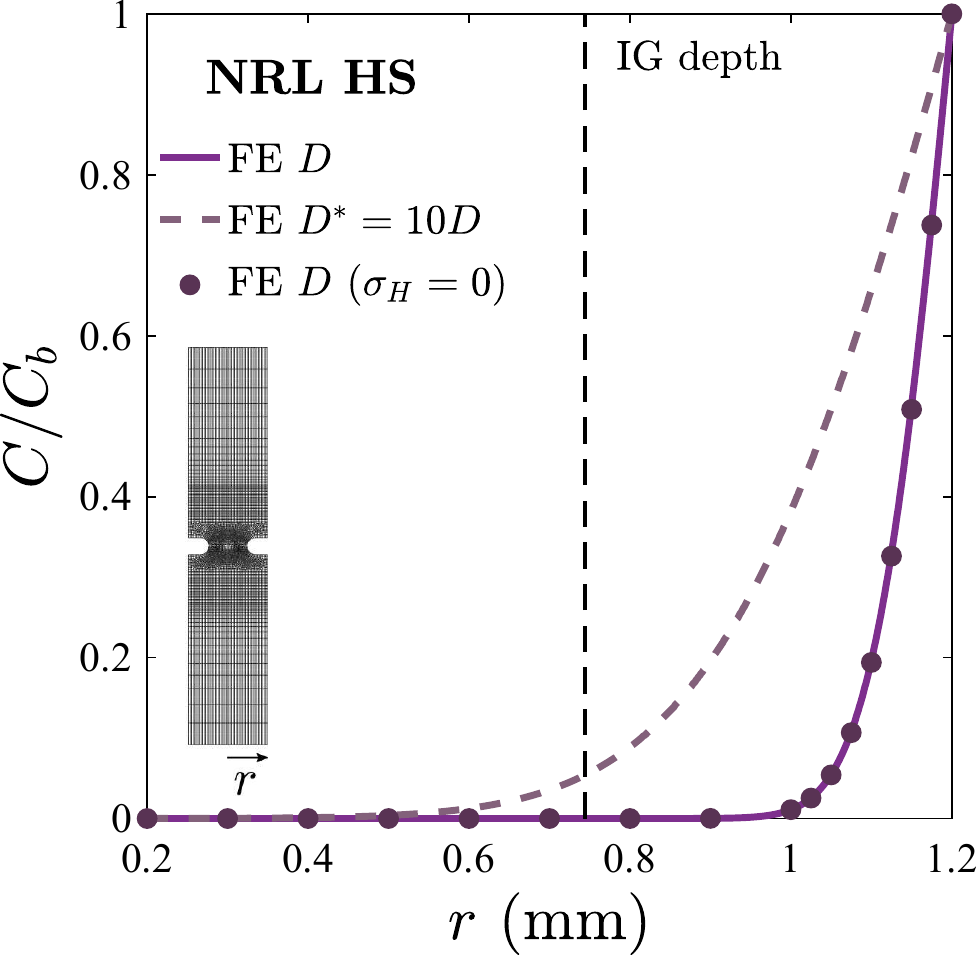}
                \caption{}
                \label{fig:NRLHSDiff}
        \end{subfigure}
        }       
        \caption{Normalized hydrogen concentration profiles for (a) Allvac, (b) TR2, (c) NRL LS, and (d) NRL HS in the absence of hydrostatic stress (circular points) and with hydrostatic stress accounted for (solid line). The dashed vertical line indicates the depth of intergranular fracture observed during testing at -1100 mV in 0.6 M NaCl. The dashed coloured curve represents the concentration profile if the hydrogen diffusivity is elevated one order of magnitude.}\label{fig:Diffusion}
\end{figure}

The simple analytical approach captured by Eq. (\ref{eq:Diffdistance}) neglects critical aspects of the experimental reality that will impact the diffusion behavior (e.g. 3D geometry and hydrostatic stress during loading \cite{Johnson2003}). As such, more rigorous finite element-based numerical analyses were performed to determine if higher levels of correspondence could be achieved by accounting for such complexities. The H concentration profiles predicted for each material heat are shown in Figure \ref{fig:Diffusion}. The results are obtained using the numerical framework described in Section \ref{Sec:FEM} but in the absence of damage, $\phi=0 \,\, \forall \, \bm{x}$ throughout the numerical experiment. Circular points denote the predictions of a purely diffusion analysis (ignoring the notch-induced stress field) while a solid line represents the results obtained with the coupled mechanical-diffusion analysis. A nearly identical concentration profile is observed if the local notch-enhanced stress profile is accounted for in the numerical simulation. The limited effect of the hydrostatic stress was not unexpected given the reduced triaxiality associated with a blunt notch, as compared to a sharp crack tip where strong effects of stress are observed on the diffusion behavior \cite{TAFM2017,IJHE2016}. In all material heats, the model results show that the H ingress is below 50\% of the intergranular fracture depth (indicated by a vertical dashed line); $\sim 30\%$ for NRL LS at -1100 mV$_{SCE}$.\\ 

The highly localized calculated concentration profile relative to the observed depth of intergranular cracking could potentially be caused by the use of a H diffusivity ($D=1.3 \times 10^{-10}$ cm$^2$/s) that is lower than reality. As summarized by Ai \textit{et al.} \cite{Ai2013}, H diffusivities ranging from $D \sim 8 \times 10^{-11}$ to $ D \sim 4 \times 10^{-10}$ cm$^2$/s have been measured in Monel K-500, depending on heat treatment and analysis methodology. In addition to such experimentally-induced variations, `fast' diffusion pathways (e.g. grain boundaries) could also facilitate deeper H penetration over expectations based on the lattice H diffusivity. Such enhanced H diffusion along grain boundaries is supported by both experimental \cite{Kimura1987,Oudriss2012,Brass1996,Tsuru1982,Mutschele1987,Harris1991,Ladna1987,Palumbo1991,Arantes1993,Doyle1995} and computational results \cite{OsmanHoch2015,Zhou2019,Zhu2019,DiStefano2015}, which collectively indicate that this diffusivity can be 1000-fold faster than the lattice H diffusivity in Ni. However, the potency of this accelerated local diffusion on the `effective' diffusivity is strongly sensitive to grain boundary character \cite{Zhou2019,Zhou2017,DiStefano2015}, grain boundary connectivity \cite{OsmanHoch2015}, grain size \cite{Brass1996,Doyle1995,Yao1991}, the presence of other solutes at the grain boundary \cite{Zhu2019}, and the H concentration \cite{Mutschele1987,Arantes1993,Zhu2019}. While significant scatter exists in the literature \cite{Tsuru1982,Harris1991,Yao1991,Yao1997}, experimental studies conducted on pure Ni with grain sizes similar to the materials used herein (10-30 $\mu$m) generally observed a 2 to 8-fold increase in the effective diffusivity \cite{Kimura1987,Brass1996,Ladna1987,Doyle1995}. An 8-fold increase in the effective diffusivity is also nominally consistent with the maximum enhancement predicted for grains larger than 1 $\mu$m by the computational results of Osman Hoch \textit{et al.} \cite{OsmanHoch2015}.\\

To assess whether such possible variations in H diffusivity could explain the discrepancy between the intergranular fracture depths and the H penetration distances, the concentration profile for each material was calculated using a H diffusivity ($D=1.3 \times 10^{-9}$ cm$^2$/s) that is an order of magnitude faster than that used in the original calculation ($D=1.3 \times 10^{-10}$ cm$^2$/s). The modeling results with this enhanced diffusivity (dashed line in Figure \ref{fig:Diffusion}) do result in a non-zero H concentration at the intergranular fracture distance.  However, the H concentrations achieved at the intergranular fracture depth are likely below the level necessary for the onset of H-induced cracking. Specifically, it is reasonable to assume (based on the results of Figure \ref{fig:IGdepth}b) that the diffusible H concentrations reported for each material heat (Table \ref{Tab:DiffusibleH}) at the most positive applied potential before intergranular fracture was observed (e.g. -850 mV$_{SCE}$ for TR2) are the minimum concentrations required for intergranular cracking. As such, H-induced intergranular fracture is not expected at -1100 mV$_{SCE}$ in the tested material heats until $C/C_b$ of at least $>0.5$ (Allvac and NRL HS), $>0.1$ (TR2), and $>0.05$ (NRL LS) are achieved. As shown by the dashed line in Figure \ref{fig:Diffusion}, even with a 10-fold enhancement in the H diffusivity, none of the tested material heats reach the H concentrations necessary for intergranular fracture at the experimentally observed intergranular fracture depth. In fact, in order for $C/C_b$ to reach a value of 0.5 at the intergranular fracture depth in the NRL HS and Allvac alloys tested at -1100 mV$_{SCE}$, simulations indicate that H diffusivities of $D=5.4 \times 10^{-9}$ and $D=7.1 \times 10^{-9}$ cm$^2$/s would be required, respectively. These values are approximately 40 and 55-fold larger than experimentally measured H diffusivity for these material heats \cite{RinconTroconis2017}, suggesting that neither experimental variability nor `fast' diffusion along grain boundaries are sufficient to induce the observed depth of intergranular fracture. This limited influence of grain boundary diffusion in the current study was not unexpected given the grain boundary characteristics of the tested material heats. Specifically, prior work on NRL HS demonstrates that the grain size and the fraction of low-angle and coincident site lattice (CSLs; predominantly $\Sigma$3) boundaries were 11.2 $\mu$m and $\sim 62\%$, respectively \cite{Harris2016}. Based on the simulations of Osman Hoch \textit{et al.}, these grain boundary characteristics would therefore result in the effective H diffusivity being $\sim 70\%$ of the true lattice diffusivity \cite{OsmanHoch2015}, implying that a grain boundary effect should not be expected in the current testing.\\

The above results suggest that H diffusion from the broad surface is insufficient to induce the observed depth of intergranular fracture in any of the material heats, even when aided by applied stresses or the expected contribution of `fast' diffusion pathways. As such, these findings support the occurrence of sub-critical crack growth during the SSRT experiment. During such growth, the electrochemical conditions responsible for H generation move along with the propagating crack, which functionally acts as a moving line source in the context of the diffusion problem \cite{Johnson1974,Gangloff2014}. This set of boundary conditions has previously been shown to enable embrittlement to depths well beyond what would be possible from diffusion from the specimen surface \cite{Johnson1974}. To validate this finding and enable commentary on its implications to the SSRT testing method, an experimentally-calibrated phase field model is utilized in the following sections to predict the onset and subsequent propagation of fracture in the absence and presence of H.

\subsection{Phase field modeling of SSRT experiments in laboratory air}
\label{Sec:ReAir}

\begin{figure}[H]
\makebox[\linewidth][c]{%
        \begin{subfigure}[b]{0.65\textwidth}
                \centering
                \includegraphics[scale=0.6]{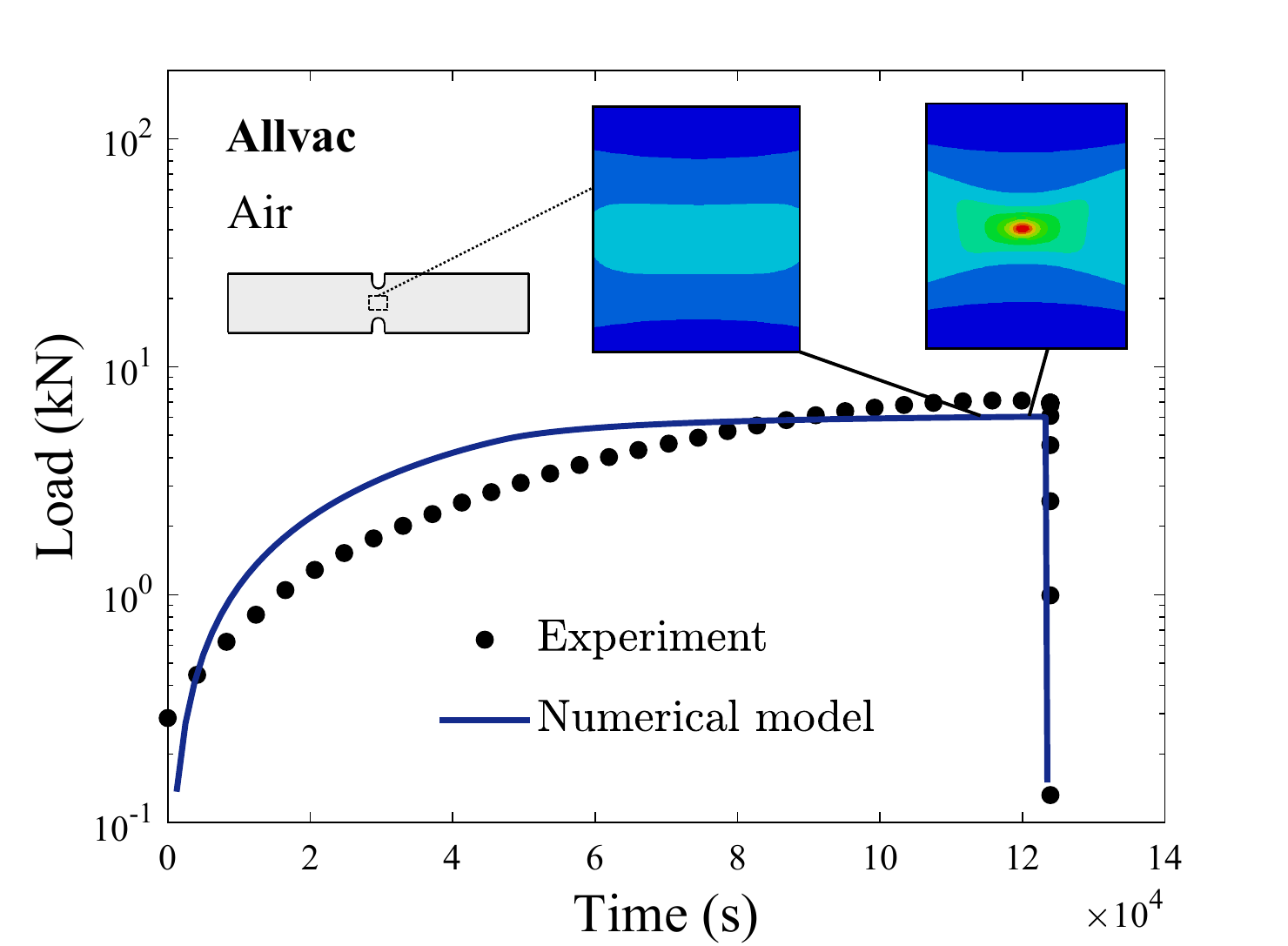}
                \caption{}
                \label{fig:AllvacAir}
        \end{subfigure}
        \begin{subfigure}[b]{0.65\textwidth}
                \raggedleft
                \includegraphics[scale=0.6]{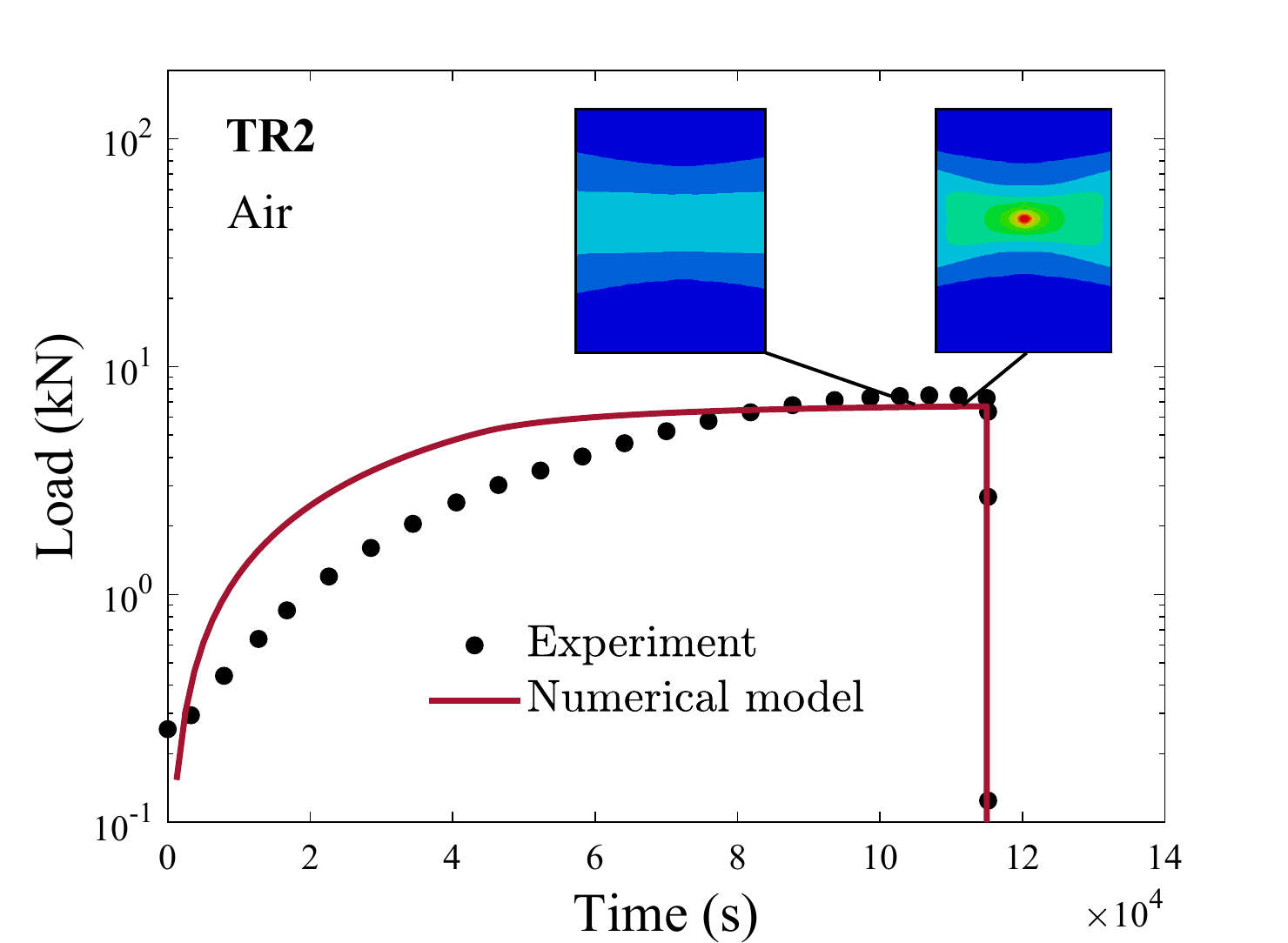}
                \caption{}
                \label{fig:TR2Air}
        \end{subfigure}}

\makebox[\linewidth][c]{%
        \begin{subfigure}[b]{0.65\textwidth}
                \centering
                \includegraphics[scale=0.6]{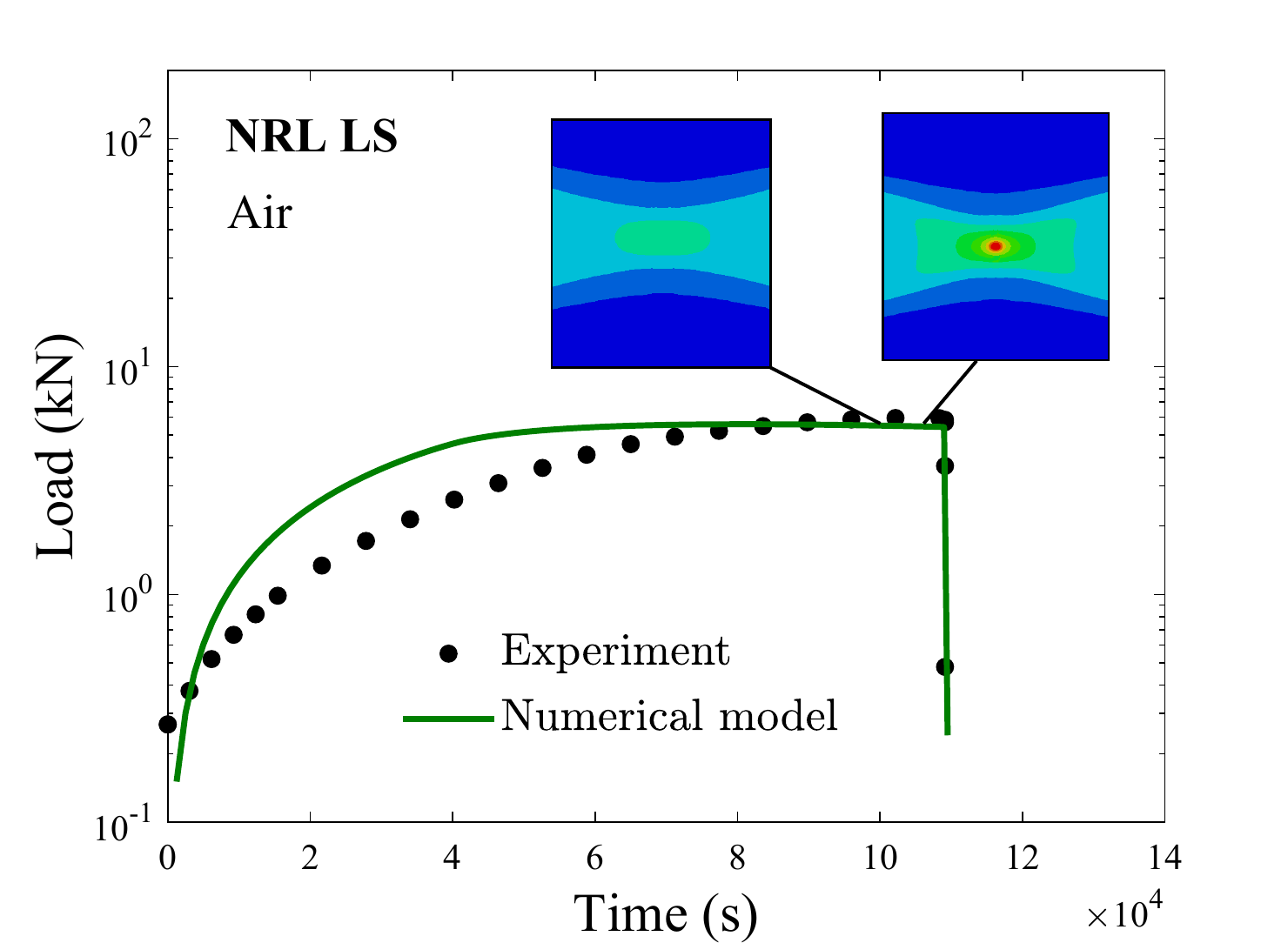}
                \caption{}
                \label{fig:NRLLSAir}
        \end{subfigure}
        \begin{subfigure}[b]{0.65\textwidth}
                \raggedleft
                \includegraphics[scale=0.6]{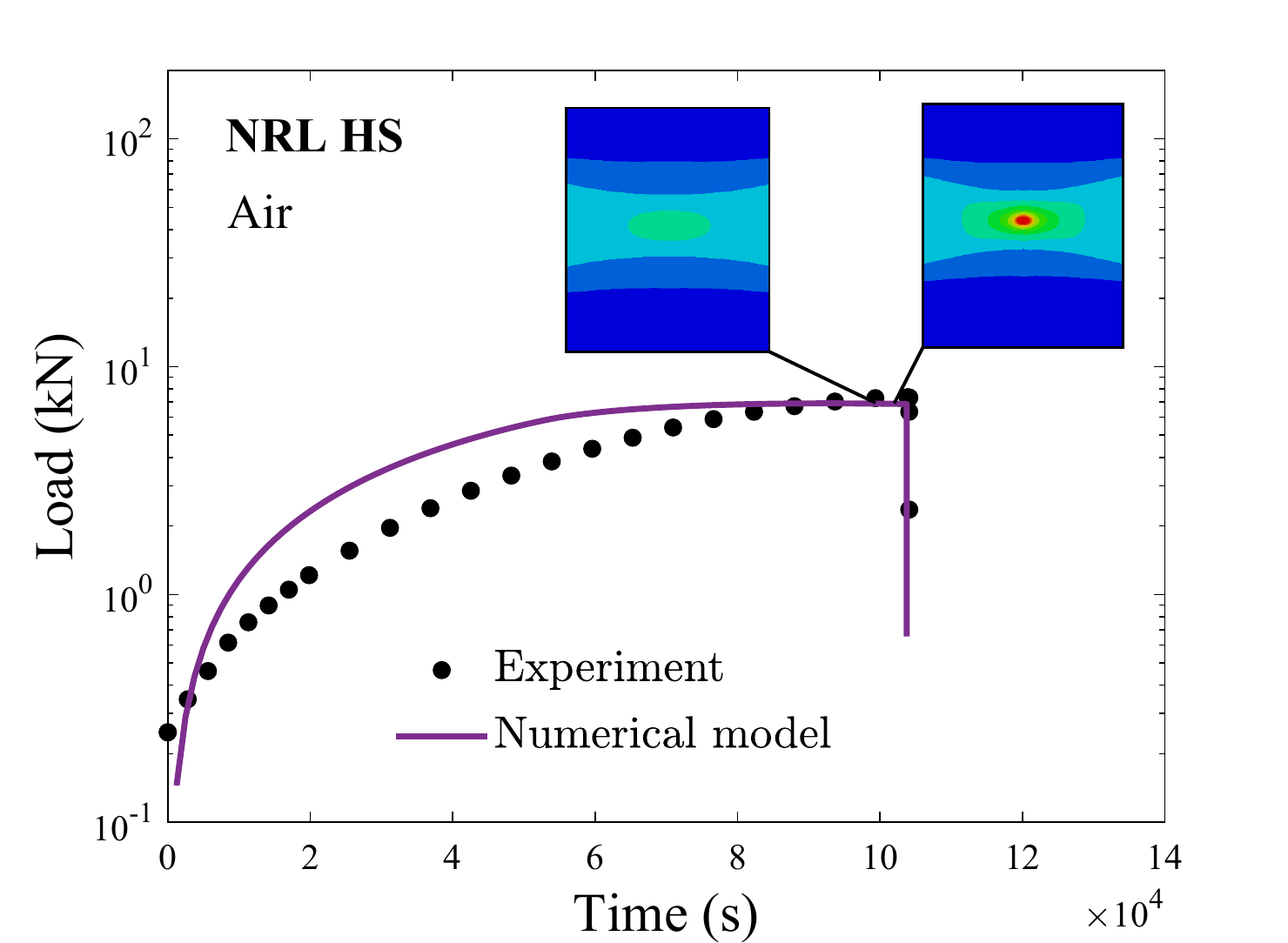}
                \caption{}
                \label{fig:NRLHSAir}
        \end{subfigure}
        }       
        \caption{Comparison of the simulated and experimental load versus time response in the absence of hydrogen for (a) Allvac, (b) TR2, (c) NRL LS, and (d) NRL HS. The inset images indicate the spatial distribution of the phase field damage parameter over a domain of a width of $\sim$0.8 mm centered at the mid-plane of the specimen. Blue and red colors correspond to the completely intact and the fully broken state of the material, respectively.}\label{fig:Air}
\end{figure}

Under the phase field modeling paradigm, the fracture response in the absence of H is characterized by the fracture energy $G_0$ and phase field length scale $\ell$. For the current simulations, $\ell$ is assumed to be 0.015 mm and the value of $G_0$ was calibrated for each material heat to match the onset of failure in the experimental force versus time data collected during testing in laboratory air (Figure \ref{fig:ExptResults}). The magnitude of $\ell$ governs the size of the fracture process zone, which is on the order of tens of microns for high strength metals in aggressive environments. One should note that the ratio $\ell/G_0$ is what governs the critical stress, see Eq. (\ref{eq:effectiveS}). Thus, the magnitude of $\ell$ is typically chosen based on mesh size considerations, approximately 7 elements are needed to resolve $\ell$ \cite{CMAME2018}, and the experimental result can then be fitted by calibrating $G_0$. Here, a total of approximately 24000 axisymmetric quadratic quadrilateral elements are employed, with the characteristic element size being $h<\ell/10$. The values of $G_0$ for each material heat that provide the best fit for these laboratory air SSRT experiments are shown in Table \ref{Tab:G0} and a comparison between the experimental results and numerical simulations using these values of $G_0$ and $\ell$ are shown in Figure \ref{fig:Air}. The inset maps depicting the spatial distribution of the phase field parameter $\phi$ in the specimen cross-section show that the model predicts that damage will initiate in the center, high-constraint region of the specimen (indicated by the red color contour). The onset of damage is very close to maximum force, as it is triggered by the necking-induced stability. Such results are consistent with expectations for a ductile fracture event in a notched specimen \cite{Needleman1984}.\\

\begin{table}[H]
\centering
\caption{Fracture energy $G_0$ obtained for each material lot by calibrating with the experiments in air.}
\label{Tab:G0}
   {\tabulinesep=1.2mm
   \begin{tabu} {cccccc}
       \hline
& Allvac & TR2 & NRL LS &  NRL HS  \\ \hline
$G_0$ (MPa$\cdot$mm) & 18.5 & 18.1 & 15.4 & 16.9 \\\hline
   \end{tabu}}
\end{table}

While Figure \ref{fig:Air} shows reasonable correspondence between the experimental results and the model, the finite element results (with and without phase field damage) over-predict the hardening behavior in each material heat. Specifically, the numerical results have an increased force for a given time relative to the experimental data. This discrepancy is not unexpected given the inherent differences between the modeling configuration and experimental set-up. The model only  accounts for the specimen response utilizing the uniaxial stress-strain data (Table \ref{Tab:MaterialProperties}) to predict the hardening behavior, with $G_0$ only informing the conditions required for the onset of fracture. Conversely, the experimental data inherently includes displacement responses associated with both the sample and the load train of the frame operated under stroke control; due to the notched geometry of the specimens (Figure \ref{fig:Specimen}), an extensometer was not employed during testing. Therefore, the displacement observed during the experiment (which is proportional to time since all testing was operated at a fixed stroke displacement rate) is described by: 
\begin{equation}\label{eq:compliance}
    \delta_T \left( F \right) = \delta_M \left( F \right) + \delta_S \left( F \right)
\end{equation}

\noindent where $\delta_T$ is the total displacement, $\delta_M$ is the machine displacement, and $\delta_S$ is the specimen displacement, all of which are functionally dependent on the applied force ($F$). A common testing artifact induced by this coupled displacement is an apparent reduction in the elastic modulus due to the dominant role of $\delta_M$  during the initial stages of deformation \cite{Davis2004}.  This coupled displacement is likely the source of the discrepancy between the experimental data and model prediction. The implication of this artifact is that the actual displacement of the test specimen ($\delta_S$) will be less than that predicted by the model, suggesting that the $G_0$ calibrated from the experimental data is an over-prediction. However, given that damage does not initiate until the $G_0$ criteria is locally satisfied, it is expected that the use of an inflated $G_0$ will yield conservative results, since an increased amount of deformation relative to reality is inherently required to induce the onset of fracture. The impact of this artifact on the interpretation of modeling results for H-charged specimens will be explored in subsequent sections.

\subsection{Phase field modeling of SSRT experiments in 0.6 M NaCl exposed to cathodic polarization}
\label{Sec:ReHcracking}

Using the experimentally-calibrated values of $G_0$ from the testing conducted in laboratory air (Table \ref{Tab:G0}), the H damage coefficient $\chi$ for each material heat was calibrated using the experimental results obtained under the most aggressive environmental conditions (0.6 M NaCl at -1100 mV$_{SCE}$; Figure \ref{fig:ExptResults}). $\chi$ establishes the extent to which $G_0$ is degraded as a function of H concentration Eq. (\ref{eq:G0X}) and is fit such that the final fracture of the model is consistent with that observed during the experiments. Variation of the values of $\chi$ (shown in Table \ref{Tab:XtoMaterialLot}) indicate a change in the H potency for the different heats of Monel K-500; this is consistent with the results of prior fracture mechanics-based testing \cite{Harris2016}.  Coupling (1) variations in $\chi$ and (2) lot-to-lot differences in the H-uptake and solubility, can be used as the basis to understand the factors governing lot specific HEAC susceptibility.  For example, despite consistently exhibiting increased resistance to H-assisted cracking \cite{RinconTroconis2017,Harris2016}, the Allvac heat was found to have a $\chi$ value (0.85) similar to the more susceptible TR2 (0.86) and NRL LS (0.82) heats. The higher levels of HEAC resistance despite a similar $\chi$ likely arises from the low concentration of H available to participate in the fracture process relative to the other heats (Table \ref{Tab:DiffusibleH}).\\

It is important to note that the value of $\chi$ will likely be affected by other modeling assumptions, such as accounting for the increase in apparent solubility due to hydrostatic stresses \cite{Turnbull1996,DiLeo2013,Diaz2016b} or the use of strain gradient plasticity to resolve micro-scale deformation \cite{AM2016,CM2017}. Moreover, physical interpretation of the values of $\chi$ is hindered by the phenomenological, macroscopic approach adopted, where both the elastic and plastic parts of the strain energy density contribute to fracture and where the damage micromechanisms are not explicitly resolved. While the phase field modeling framework can be extended to provide a direct connection with potential HE mechanisms (AIDE \cite{Lynch1988}, HELP \cite{Birnbaum1994}, HEDE \cite{Oriani1972,Troiano2016}, etc.), our objective is to provide a hydrogen-assisted fracture mechanics assessment of cracking in SSRT. Thus, $\chi$ does not have the same physical meaning as the hydrogen damage coefficients that can be determined from the atomistic simulations of Jiang and Carter \cite{Jiang2004a} or Alvaro \textit{et al.} \cite{Alvaro2015}. In these prior studies, a brittle fracture paradigm was invoked, where $\chi$ would generally be interpreted as the potency of hydrogen in reducing the ideal work of fracture. However, a \textit{total} energy framework is here employed \cite{Jokl1980}, given that Monel K-500 exhibits tangible plastic deformation prior to failing via intergranular fracture, even when charged with hydrogen contents reaching $>$200 weight part per million \cite{Harris2020}.\\

\begin{table}[H]
\centering
\caption{Sensitivity of the hydrogen damage coefficient $\chi$ to the material heat for $E_A=-1100$ mV.}
\label{Tab:XtoMaterialLot}
   {\tabulinesep=1.2mm
   \begin{tabu} {ccccc}
       \hline
 Material lot & Allvac & TR2 & NRL LS & NRL HS  \\ \hline
 Hydrogen damage coefficient $\chi$  & 0.85 & 0.86 & 0.82 & 0.79 \\\hline
   \end{tabu}}
\end{table}

\begin{figure}[H]
\makebox[\linewidth][c]{%
        \begin{subfigure}[b]{0.65\textwidth}
                \centering
                \includegraphics[scale=0.6]{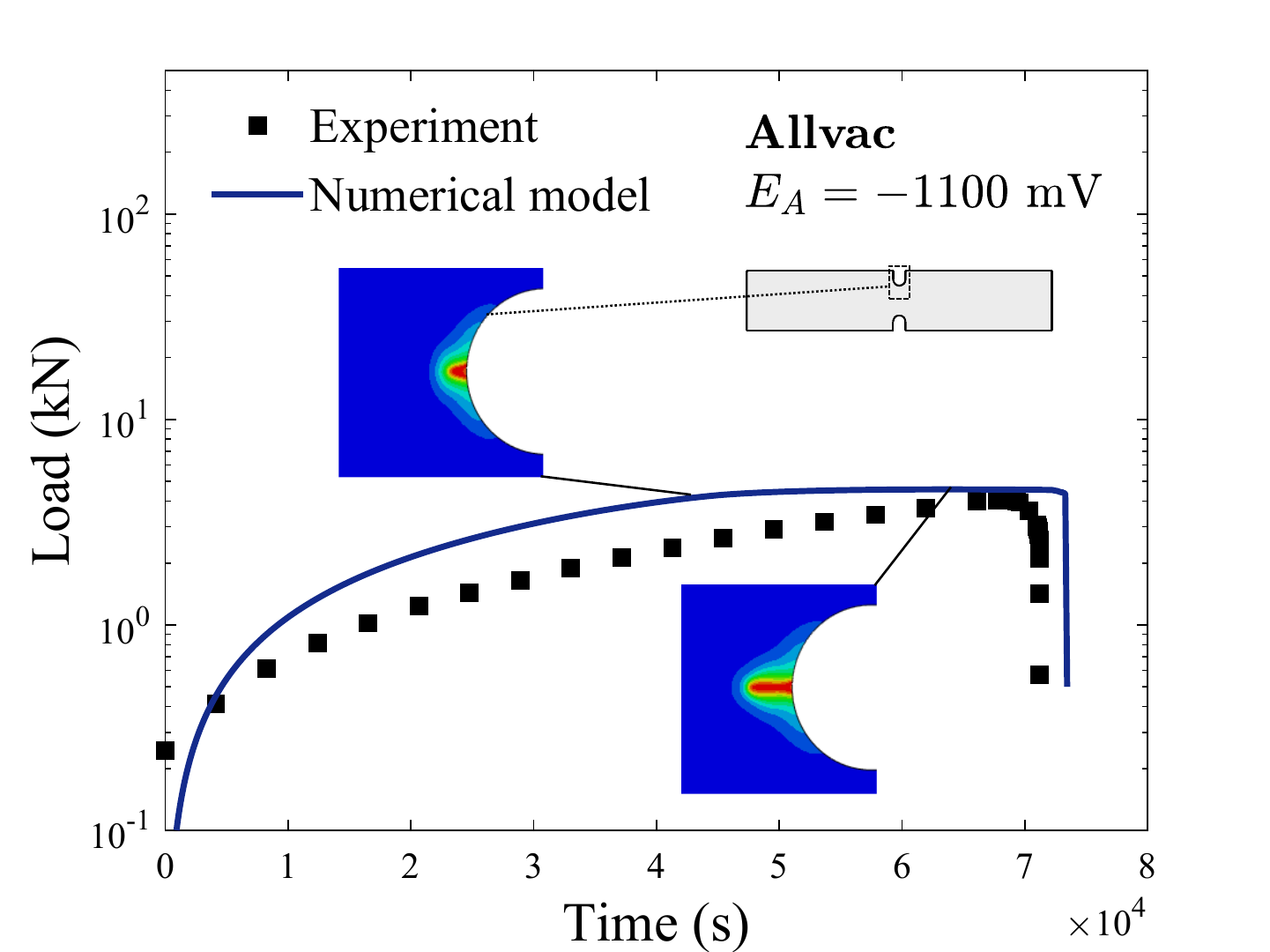}
                \caption{}
                \label{fig:AllvacH}
        \end{subfigure}
        \begin{subfigure}[b]{0.65\textwidth}
                \raggedleft
                \includegraphics[scale=0.6]{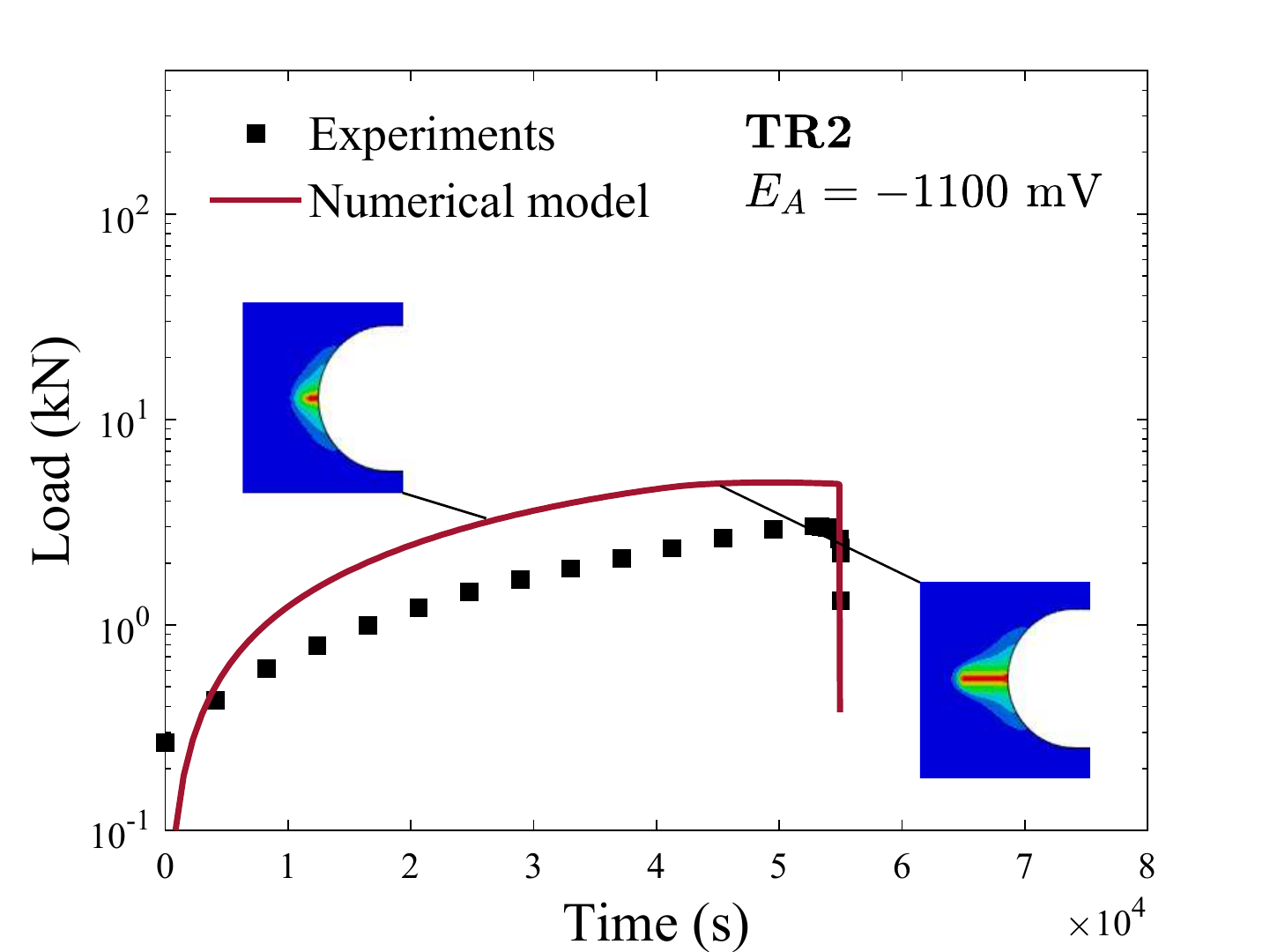}
                \caption{}
                \label{fig:TR2H}
        \end{subfigure}}

\makebox[\linewidth][c]{%
        \begin{subfigure}[b]{0.65\textwidth}
                \centering
                \includegraphics[scale=0.6]{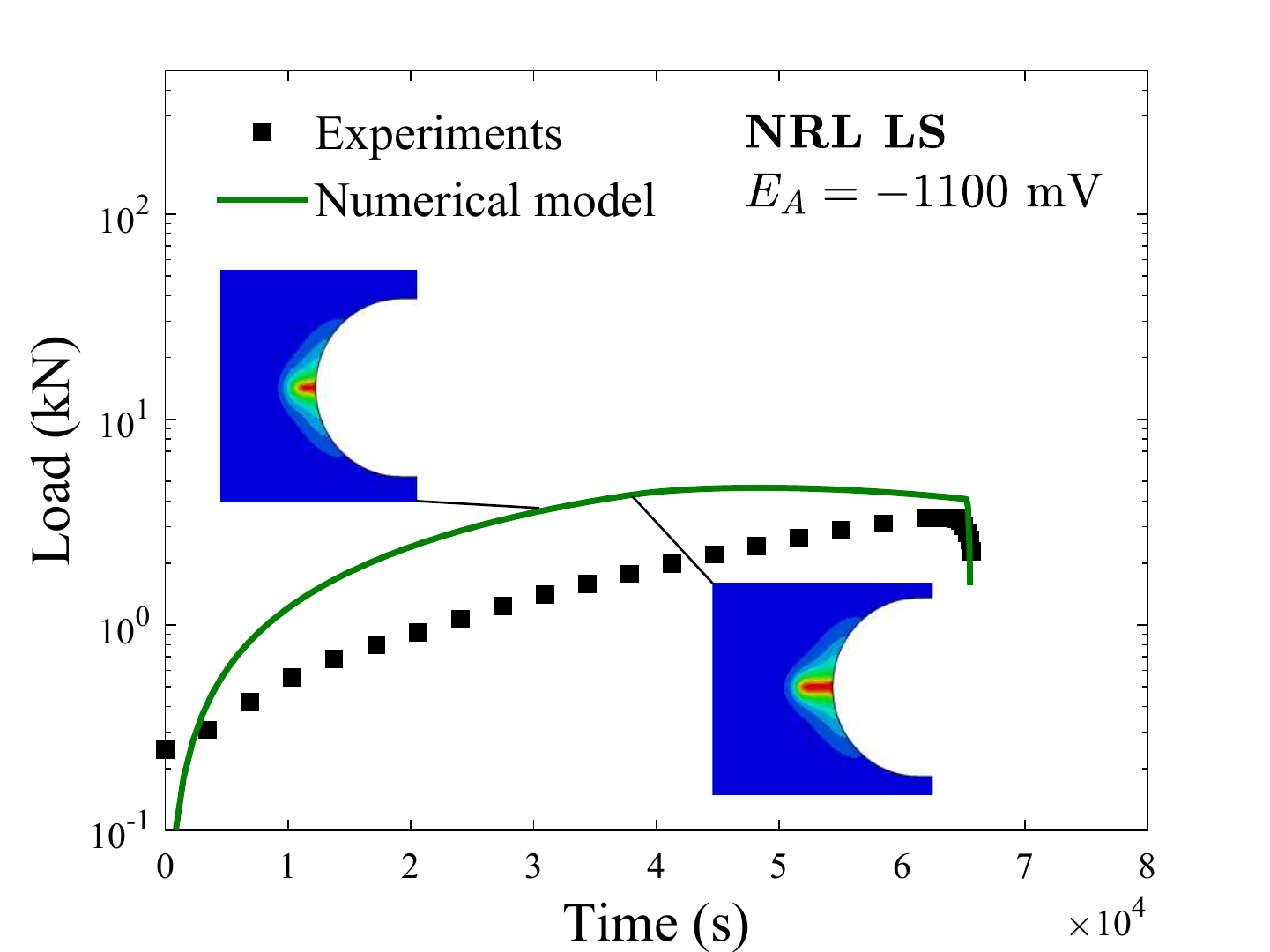}
                \caption{}
                \label{fig:NRLLSH}
        \end{subfigure}
        \begin{subfigure}[b]{0.65\textwidth}
                \raggedleft
                \includegraphics[scale=0.6]{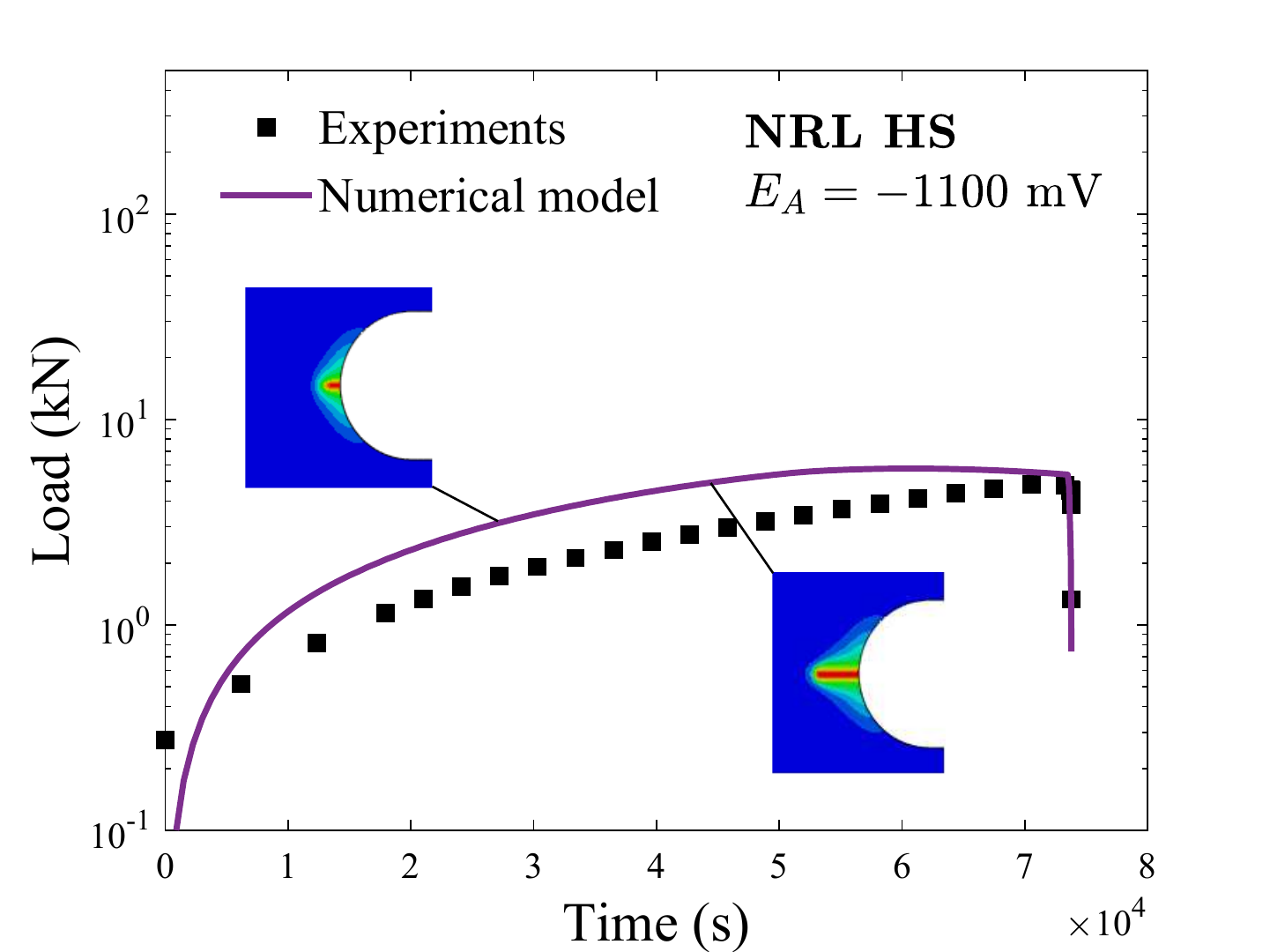}
                \caption{}
                \label{fig:NRLHSH}
        \end{subfigure}
        }       
        \caption{Comparison of the simulated and experimental load versus time response at -1100 mVSCE in 0.6 M NaCl for (a) Allvac, (b) TR2, (c) NRL LS, and (d) NRL HS. The inset images indicate the spatial distribution of the phase field damage parameter over a domain of a width of $\sim$1.2 mm. Blue and red colors correspond to the completely intact and the fully broken state of the material, respectively.}\label{fig:H}
\end{figure}

The experimental and simulated force versus time curves for each material heat at -1100 mV$_{SCE}$ in 0.6 M NaCl are shown in Figure \ref{fig:H}. A difference in the hardening behavior between the model prediction and experimental data is noted for the testing in 0.6 M NaCl at –1100 mV$_{SCE}$. However, as was argued for the similar discrepancy in the lab air results, this difference is reasonably attributed to the inclusion of both the specimen and actuator displacement in the experimental data (\ref{eq:compliance}).  The results demonstrate reasonable agreement between the predicted and experimentally-observed rapid decrease in load associated with specimen failure. Furthermore, the inset maps of the spatial distribution of the phase field parameter $\phi$ in the specimen cross-section (Figure \ref{fig:H}) reveal the initiation of cracks at the root of the notch in all material heats. This location is notably different from the initiation of damage at the center of the SSRT specimen in the laboratory air testing (Figure \ref{fig:Air}). Interestingly, the model predicts crack initiation at between 40 to 60\% of the total time to failure for the SSRT experiments at -1100 mV$_{SCE}$. Critically, these results imply that sub-critical crack growth will occur for 40 to 60\% of the test duration. As such, the modeling results provide firm support for the preceding diffusion-based arguments that suggest sub-critical crack growth is required to obtained the experimentally-measured intergranular fracture depths (Figure \ref{fig:IGdepth}). \\

\begin{figure}[H]
\centering
\includegraphics[scale=0.9]{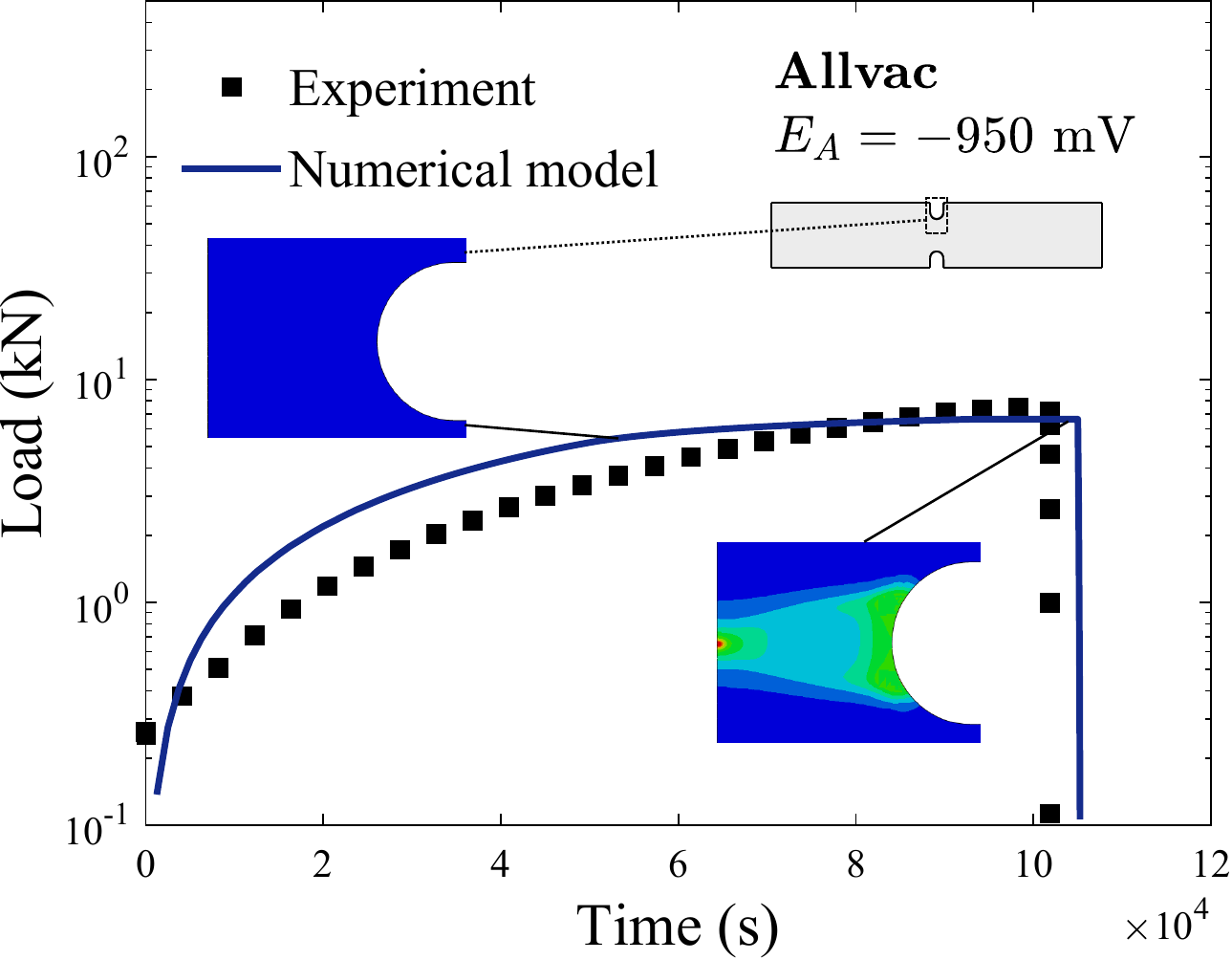}
\caption{Comparison of the simulated and experimental load versus time response at -950 mV$_{SCE}$ in 0.6 M NaCl for the material heat. The inset images for each material heat indicate the spatial distribution of the phase field damage parameter in the cross-section of the SSRT specimen, with the blue representing $\phi=0$ and red representing $\phi=1$.}
\label{fig:FvsTimeAllvac950}
\end{figure}

It is possible that this artifact could influence the calibrated value of $\chi$, thereby complicating the interpretation of the modeling results. The validity of the modeling parameters is tested by using the calibrated values of $\chi$ and $G_0$ for each material heat to model SSRT experiments conducted at intermediate applied potentials (e.g. –950 or –850 mV$_{SCE}$). If $\chi$ is independent of (or insensitive to) the coupled displacements artifact, then it should reasonably reproduce the rapid decrease in applied load at the onset of final failure for the intermediate applied potentials. The modeling results for Allvac at -950 mV$_{SCE}$ (Figure \ref{fig:FvsTimeAllvac950}) show a similar level of agreement with the experimental data as was observed for Allvac at -1100 mV$_{SCE}$. Such results were also observed for the other material heats at intermediate applied potentials.\footnote{Note that the tests conducted on NRL HS and NRL LS at -850 mV$_{SCE}$ were not modeled due to their increased time to failure relative to laboratory air testing, which is speculatively attributed to test-to-test scatter in mechanical properties.} These results provide confidence in the general trends predicted by the model (i.e. the initiation of sub-critical cracking well before final fracture at -1100 mV$_{SCE}$). Interestingly, the model also correctly predicts that at -950 mV$_{SCE}$ the final fracture will initiate at the center of the specimen, as demonstrated by the inset maps of $\phi$ in the specimen cross-section. However, damage proximate to the notch root (indicated by the green field in the inset map of Figure \ref{fig:FvsTimeAllvac950}) is also observed. Speculatively, this H-induced surface damage results in the decreased time to failure for -950 mV$_{SCE}$ by degrading the resistance of the cross-section to deformation, leading to an earlier plastic instability and concomitant initiation of fracture at the center of the specimen.

\section{Discussion}
\label{Sec:Discussion}

\subsection{Evidence for the onset of sub-critical cracking during SSRT experiments}

A holistic evaluation of the presented experimental observations and modeling results provides strong evidence for the onset of sub-critical crack growth during notched SSRT experiments. First, a systematic comparison of the expected H diffusion ingress and the observed intergranular fracture depth was conducted for four different material heats of Monel K-500 tested in 0.6 M NaCl at various applied potentials. The results of this evaluation demonstrate that H diffusion from the specimen surface cannot reasonably explain the observed depth of intergranular fracture. Specifically, even when the effect of applied stress and/or an order of magnitude faster diffusivity are invoked (to simulate an effect of grain boundary diffusion), H ingress from the specimen surface is insufficient to induce the measured extent of intergranular fracture (Figure \ref{fig:Diffusion}). These results strongly support the occurrence of sub-critical crack growth during the SSRT experiments, where a moving line source phenomenon would enable the extended intergranular fracture depths \cite{Johnson1974}. To test this hypothesis, a new phase field formulation that coupled H diffusion and elastic-plastic deformation was employed. Using the experimental results from laboratory air testing and 0.6 M NaCl at -1100 mV$_{SCE}$, the critical total fracture energy $G_0$ and H damage coefficient $\chi$ were calibrated for the four material heats (Figures \ref{fig:Air}-\ref{fig:H}). These parameters were subsequently validated via modeling of tests conducted at intermediate applied potentials (-950 and -850 mV$_{SCE}$ in 0.6 M NaCl), which showed reasonable agreement for the onset of failure between the model predictions and experimental results without any modification of the calibrated $G_0$ and $\chi$ values (Figure \ref{fig:FvsTimeAllvac950}). Critically, through an evaluation of the phase field parameter $\phi$, the fracture initiation location for each material heat/environment combination can be ascertained. For the laboratory air testing that exhibited ductile microvoid coalescence as the dominant fracture morphology, fracture was found to initiate in the center of the specimen (Figure \ref{fig:Air}), consistent with expectations for notched tensile bars \cite{Needleman1984}. Conversely, modeling of the experiments at -1100 mV$_{SCE}$ that exhibited intergranular fracture revealed the initiation and subsequent propagation to failure of a crack at the notch root. Critically, and in direct agreement with the inferences drawn from the diffusion-based analysis, the cracks were found to initiate well before final fracture (i.e. sub-critical cracking), at times corresponding to 40-60\% of the time to failure. Taken together, these two results strongly suggest that sub-critical cracking can occur during SSRT experiments.\\

This conclusion is consistent with a broad literature survey of prior SSRT experimental results. For example, Holroyd and coworkers have documented the existence of sub-critical cracks during SSRT experiments on 5xxx-series Al alloys \cite{HenryHolroyd2019,Holroyd2017,Seifi2016}. In fact, through systematic x-ray computed tomography (XCT), these authors have gone so far as to estimate stress intensity solutions for the observed crack morphologies as well as attempt to characterize crack growth rates during SSRT testing \cite{HenryHolroyd2019}. Similar findings were also reported by Sampath \textit{et al.}, who utilized an imaging technique to assess the topography of matching fracture surfaces in H-exposed Ni-based alloys to determine that cracks could initiate at stresses as low as 33\% of the fracture stress \cite{Sampath2018}. M\'athis \textit{et al.} leveraged acoustic emission to assess the efficacy of SSRT experiments on stainless steel tubulars, which identified acoustic events associated with crack initiation and propagation well before final fracture, with subsequent scanning electron microscopy confirming the presence of stress corrosion cracks in the test specimen \cite{Mathis2011}. SSRT experiments by Haruna and Shibata on cylindrical 316L stainless steel exposed to aqueous NaCl solutions in which the tests were interrupted and then cross-sectioned for evaluation revealed crack initiation and propagation prior to failure \cite{Haruna1994}. Lee \textit{et al.} speculated that the fracture of AA2024-T351 SSRT specimens exposed to aqueous chloride solutions resulted from the formation of pits, which enabled the initiation and propagation of cracks, leading to final failure \cite{Lee2012}. Lastly, modeling of SSRT experiments by Garud explicitly includes the formation and propagation of `microcracks' as the means by which the net-section load-carrying capacity is diminished \cite{Garud1990}. The current diffusion and phase field modeling results support these literature reports that demonstrate sub-critical crack growth can occur during SSRT experiments. As discussed by Ahluwalia \cite{Ahluwalia1993}, the onset of such cracking can alter the assessment of SSRT results by complicating the interpretation of typical metrics like time-to-failure, breaking stress, and ductility. The computation tools developed herein enable an efficient interrogation of the effect that various environment, loading, and material combinations may have on the accuracy and validity of measured SSRT metrics. In the following section, the implications of such complications on the evaluation of the time-to-failure metric is explored leveraging the results of the phase field modeling shown in Figure \ref{fig:H}.

\subsection{Implications for the interpretation of SSRT experiments}
\begin{figure}[H]
\centering
\includegraphics[scale=0.7]{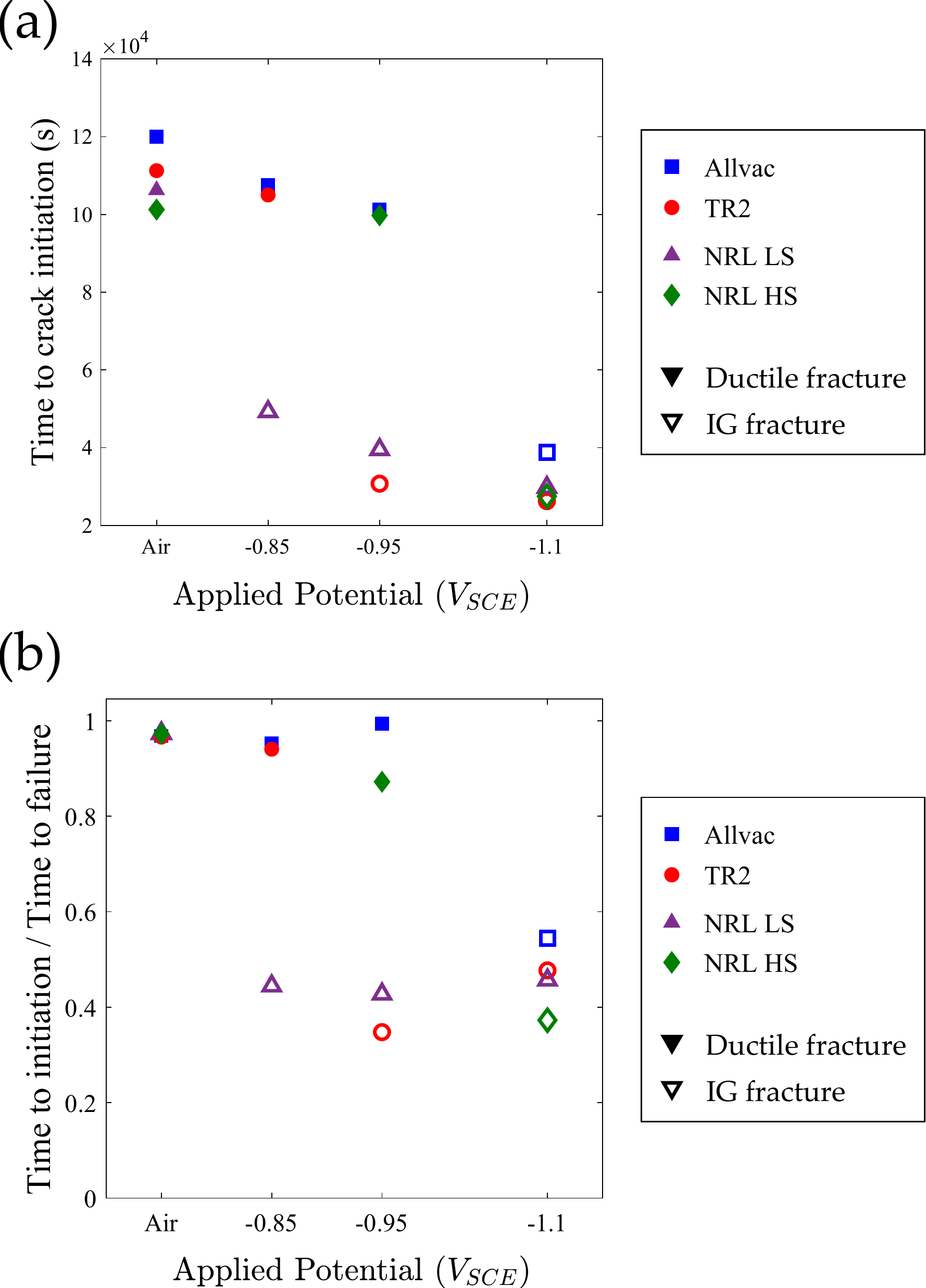}
\caption{(a) Time for crack initiation and (b) ratio of the time for crack initiation and time for specimen failure as a function of applied potential for the four tested material heats of Monel K-500. Open and closed symbols indicate specimens that exhibited intergranular (IG) and ductile fracture, respectively.}
\label{fig:TimeVSEa}
\end{figure}

As demonstrated in Figure \ref{fig:H}, the phase field model enables the assessment of the time for crack initiation, which provides an opportunity to qualitatively establish the implications of sub-critical crack growth on the interpretation of SSRT metrics. As shown in Figure \ref{fig:TimeVSEa}a, the model-predicted time to crack initiation is found to decrease with increasingly more negative applied potentials for all material heats; prior work in Monel K-500 demonstrated that such behavior is due to increasing H generation with increasingly negative applied potential \cite{RinconTroconis2017,Harris2016,Gangloff2014,Burns2016a}. However, as the most aggressive potential is reached, there is a significant and important difference between the time for crack initiation and the time for failure data. Specifically, the ranking of the HEAC susceptibility of the materials heats based on their time to failure (a common SSRT metric \cite{Henthorne2016}) is not consistent with the ranking obtained when using the time to crack initiation (which is more relevant to the true HEAC susceptibility). For example, despite being considered one of the more HEAC-resistant heats of Monel K-500 \cite{RinconTroconis2017,Harris2016} based on the traditional SSRT metrics, NRL HS is found to exhibit a similar time to crack initiation as the EAC-susceptible NRL LS and TR2 material heats. Such differences further underscore the reality that the SSRT experiment is a composite of the times required for crack incubation, initiation, Stage I propagation, Stage II propagation, and final fracture. The interpretation of the SSRT test becomes even more convoluted if one considers the ratio of the time to initiation and time to failure, shown in Figure \ref{fig:TimeVSEa}b. Under this paradigm, NRL HS would be considered the worst-performing material, as cracking is found to initiate at times of less than 40\% of the time to failure. 

As discussed by Ahulwalia \cite{Ahluwalia1993}, this ambiguity in interpreting the results of the SSRT experiments (e.g. three different conclusions are drawn if the data are evaluated in three different ways) highlights the complications introduced upon the onset of sub-critical cracking. Such difficulties also extend to the use of stress-based metrics (e.g. breaking stress), where it would be necessary to determine whether (1) the remote stress at crack initiation, (2) the remote stress at fracture, or (3) a calculated stress intensity is the most rigorous metric. Coupling this challenge in interpretation with the explicit caveat in ASTM Standard G129 that SSRT ``results are not intended to necessarily represent service performance'' \cite{ASTMG129}, raise questions regarding the efficacy of the SSRT methodology to reliably assess/rank H embrittlement susceptibility (and EAC susceptibility in general). The SSRT experimental approach has several attractive elements relative to other testing methodologies (e.g. low-cost, efficient, modular, simple to perform, etc. \cite{Kim1979,Henthorne2016,McIntyre1988,Parkins1979}). Moreover, it is likely that there are environment/material/loading combinations for which the SSRT remains a useful screening approach. For example, if H diffusion is sufficiently fast to enable the attainment of an uniform H concentration across the SSRT specimen diameter and the alloy toughness is low (\textit{e.g.} high-strength steels \cite{Lee2007}), then it is possible that final fracture could occur immediately upon crack initiation. However, such bounding conditions have not been firmly established to-date, underscoring the need for additional studies to assess the conditions under which the SSRT methodology represents a reasonable pathway for assessing/screening for EAC susceptibility.\\

Based on the current work, several initial suggestions can be made. First, the use of notched tensile specimens for SSRT should be carefully considered prior to use, as these specimens appear to be particularly susceptible to the onset of sub-critical cracking due to the stress concentration present at the notch root. While the use of such notches enables closer approximation to a typical bolt geometry, ensures fracture occurs at a specific location, and induces failure at reduced loads, these benefits may be out-weighed by the complications associated with sub-critical cracking. Second, the current study establishes the utility of a diffusion-based analysis framework to ascertain whether or not sub-critical cracking occurred during SSRT experiments in H-containing/producing environments. Given the simple geometries generally employed for the majority of SSRT experiments \cite{Henthorne2016}, the diffusion problem can be readily solved either analytically \cite{Shewmon2016} or with the aid of finite element software packages. Comparison of these calculations with the depth of EAC-induced brittle fracture provides a rigorous pathway for assessing the validity of SSRT experiments. Simply put, if H cannot reasonably diffuse to the observed extent of brittle fracture, then it is possible that sub-critical cracking occurred and additional efforts should be made before interpreting obtained SSRT metrics. Lastly, if secondary cracks are observed along the gage length of the SSRT specimen after the conclusion of testing, it is possible that sub-critical cracking may have occurred. To assess this possibility, the interrupted testing and cross-sectioning methodology of Haruna and Shibata \cite{Haruna1994} could be used to confirm that such cracks do not form until final fracture. \\ 

The employed modeling framework offers a unique tool to assess the efficacy of the SSRT methodology to evaluate susceptibility to H embrittlement. Confidence in the modeling is obtained by consistency with prior literature conclusions and diffusion-based arguments as well as the ability to reasonably capture the experimental data from intermediate applied potentials (Figure \ref{fig:FvsTimeAllvac950}). However, it should be noted that the model-predicted metrics (i.e. time to crack initiation) have yet to be explicitly validated against experimental data. Towards this end, it would be useful to perform a series of systematically interrupted SSRT experiments, which would then be sectioned to experimentally evaluate when crack initiation occurs \cite{Harris2018,Haruna1994}. Alternatively, the XCT approach previously utilized by Holroyd \textit{et al.} \cite{Holroyd2017} or the imaging analysis employed by Sampath \cite{Sampath2018} could be leveraged to determine the onset of subcritical crack growth. 

\section{Conclusions}
\label{Sec:Concluding remarks}
The potential for the onset of sub-critical crack growth during slow strain rate tensile testing of Monel K-500 immersed in 0.6 M NaCl and exposed to applied potentials ranging from -850 to -1100 mV$_{SCE}$ was systematically evaluated using both H diffusion-based analyses and a new phase field formulation for H-assisted fracture. These efforts, coupled with an examination of prior literature results, reveals several important insights regarding the onset of sub-critical crack growth during SSRT experiments to assess H embrittlement susceptibility:\\

1) A systematic comparison of the observed depth of intergranular fracture and the calculated H ingress distance revealed that H cannot reasonably reach the depth of intergranular fracture. These results strongly suggest that sub-critical crack growth occurred during the SSRT experiments.\\

2) A phase field formulation that couples elastic-plastic deformation with H diffusion was developed to describe H-assisted fracture. Model parameters were calibrated using the experimental data obtained during SSRT experiments conducted in laboratory air ($G_0$) and in 0.6 M NaCl at -1100 mV$_{SCE}$ ($\chi$), and then validated via a systematic comparison with results from SSRT experiments conducted at lower overpotentials.\\
	
3) The phase field model reliably predicted the expected initiation location for fracture for all experiments: at the center of the specimen for non-embrittling environments and at the notch tip for specimens exhibiting intergranular fracture.\\
	
4) The calibrated model predicted the initiation of cracks at the notch tip at 40-60\% of the time to failure, depending on material heat. Critically, the model captured the observed degradation in time-to-failure for experiments (e.g. Allvac at -950 mV$_{SCE}$) conducted at applied potentials that did not exhibit intergranular fracture without any modification of calibrated parameters.\\
	
5) The potential impact of sub-critical cracking on the validity of reported metrics from SSRT experiments was explored via a comparison the experimentally-obtained time to failure and the model-predicted time to crack initiation. This gedankan experiment revealed a modification in the relative ordering of H embrittlement susceptibility in four material heats of Monel K-500, underscoring the important impact that sub-critical cracking can have on the interpretation of SSRT metrics.

\section{Acknowledgments}
\label{Acknowledge of funding}
Helpful discussions with Prof. Richard Gangloff at the University of Virginia are gratefully acknowledged. E. Mart\'{\i}nez-Pa\~neda acknowledges financial support from the Ministry of Economy and Competitiveness of Spain through grant MAT2014-58738-C3 and the People Programme (Marie Curie Actions) of the European Union’s Seventh Framework Programme (FP7/2007-2013) under REA grant agreement n$^{\circ}$ 609405 (COFUNDPostdocDTU). Z.D. Harris gratefully acknowledges the support of the ALCOA Graduate Fellowship administered by the School of Engineering and Applied Science at the University of Virginia. 

\section{Data Availability Statement}
The data generated during this study will be made available upon reasonable request.


\appendix
\section{Finite element implementation}
\label{App:FE}

The finite element (FE) method is used to solve the coupled mechanical-diffusion-phase field problem. Using Voigt notation, the nodal values of the displacements, phase field and H concentration are interpolated as follows,
\begin{equation}\label{Eq:Discretization}
    \bm{u}=\sum_{i=1}^m \bm{N}_i \bm{u}_i \, , \,\,\,\,\,\,\,\,\,  \phi=\sum_{i=1}^m N_i \phi_i \, , \,\,\,\,\,\,\,\,\,  C=\sum_{i=1}^m N_i C_i
\end{equation}

\noindent where $m$ is the number of nodes and $\bm{N}_i$ are the interpolation matrices - diagonal matrices with the nodal shape functions $N_i$ as components. Accordingly, the corresponding gradient quantities can be discretized by,
\begin{equation}
    \bm{\varepsilon}=\sum_{i=1}^m \bm{B}^u_i \bm{u}_i \, , \,\,\,\,\,\,\,\,\,  \nabla \phi=\sum_{i=1}^m \bm{B}_i \phi_i \, , \,\,\,\,\,\,\,\,\,  \nabla C=\sum_{i=1}^m \bm{B}_i C_i
\end{equation}

\noindent Here, $\bm{B}_i$ are vectors with the spatial derivatives of the shape functions and $\bm{B}^u_i$ denotes the standard strain-displacement matrices.

\subsection{FE discretization of the deformation-phase field problem}

The implementation of the coupling between elastic-plastic deformation and phase field fracture builds upon the work of Miehe and co-workers \cite{Miehe2010a}. Thus, nodal values of displacement and phase field order parameter are obtained by means of a staggered approach, and a history variable $H$ is introduced to ensure irreversibility of the crack phase field evolution
\begin{equation} \label{eq:history}
    H =
    \begin{cases}
      \psi  & \quad \text{if} \hspace{2mm} \psi > H_{t} \\
      H_{t}  & \quad \text{otherwise}\\
    \end{cases}
    \centering
\end{equation}
\noindent Here, \(H_{t}\) is the previously calculated energy at time increment \(t\). Thus, the history field satisfies the Kuhn-Tucker conditions.\\

Making use of the finite element discretization outlined above and considering that Eq. (\ref{Eq:weak}) must hold for arbitrary values of $\delta \bm{u}$, the discrete equation corresponding to the equilibrium condition can be expressed as the following residual with respect to the displacement field,
\begin{equation} \label{eq:residualStagU}
    \bm{r}_{i}^u =\int_\Omega \left\{ \left[(1-\phi)^{2}+k\right] {(\bm{B}_{i}^u)}^{T} \bm{\sigma} \right\} \, \mathrm{d}V
\end{equation}

\noindent with $k$ being a sufficiently small numerical parameter introduce to keep the system of equations well-conditioned; a value of $k=1 \times 10^{-7}$ is used throughout this work. Similarly, the out-of-balance force residual with respect to the evolution of the crack phase field is obtained by discretizing Eq. (\ref{Eq:weak}) and considering Eq. (\ref{eq:history}),
\begin{equation}
    r_{i}^{\phi}= \int_\Omega \left[ -2(1-\phi) N_{i} \, H + G_0 \left( C \right) \left(\dfrac{\phi}{\ell} N_{i}  + \ell \bm{B}_{i}^T \nabla \phi \right) \right] \, \mathrm{d}V
\end{equation}

\noindent The components of the consistent stiffness matrices can be obtained by differentiating the residuals with respect to the incremental nodal variables:
\begin{equation}\label{Eq:Ku}
    \bm{K}_{ij}^{\bm{u}} = \frac{\partial \bm{r}_{i}^{\bm{u}}}{\partial \bm{u}_{j}} = \int_\Omega \left[(1-\phi)^2+ k\right] {(\bm{B}_i^{\bm{u}})}^T \bm{C}_{ep} \bm{B}_j^{\bm{u}} \, \mathrm{d}V  
\end{equation}
\begin{equation}
    \bm{K}_{ij}^\phi = \dfrac{\partial r_{i}^{\phi}}{\partial \phi_{j}} = \int_\Omega \left[ \left( 2H + \dfrac{G_0 \left( C \right)}{\ell} \right) N_{i} N_{j} + G_0 \left( C \right) \ell \bm{B}_i^T \bm{B}_j \right] \, \mathrm{d}V    
\end{equation}

\noindent where $\bm{C}_{ep}$ is the elastic-plastic consistent material Jacobian.

\subsection{FE discretization of mass transport}

We assume no external flux in the surface $\partial \Omega_q$ and obtain a residual vector for the mass diffusion problem by discretizing Eq. (\ref{Eq:WeakC}) for any arbitrary virtual variation of the H concentration $\delta C$. Accordingly, the residual right hand side vector $r^C_i$ reads
\begin{equation}\label{eq:ResH}
     r_{i}^C= \int_\Omega \left[ N_i^T \left( \frac{1}{D} \frac{dC}{dt} \right) + \bm{B}_i^T  \nabla C  - \bm{B}_i^T \left( \frac{\bar{V}_H C}{RT}  \nabla \sigma_H \right)  \right] \, \mathrm{d}V 
\end{equation}

\noindent From which a diffusivity matrix can be readily defined,
\begin{equation}\label{eq:KH}
    \bm{K}_{ij}^C = \int_\Omega \left( \bm{B}_i^T  \bm{B}_j  - \bm{B}_i^T \frac{\bar{V}_H}{RT}   \nabla \sigma_H N_j   \right) \, \mathrm{d}V    
\end{equation}

\noindent where the discretization given in Eq. (\ref{Eq:Discretization}) has also been employed to interpolate the time derivatives of the nodal concentrations. The diffusivity matrix is affected by the gradient of the hydrostatic stress, $\sigma_H$, which is computed at the integration points from the nodal displacements, extrapolated to the nodes by means of the shape functions, and subsequently multiplied by $\bm{B}_i$ to compute $\nabla \sigma_H$. Second order elements are employed in the computational implementation.\\

At the same time, one can readily identify a concentration capacity matrix,
\begin{equation}
    \bm{M}_{ij} = \int_\Omega N_i^T \frac{1}{D} N_j \, \mathrm{d}V        
\end{equation}

\noindent and the discretized H transport equation reads,
\begin{equation}
    \bm{K}^C \bm{C} + \bm{M} \dot{\bm{C}} = \bm{0}
\end{equation}

\subsection{Coupled scheme}

The mechanical deformation, mass diffusion and phase field fracture problems are weakly coupled. First, elastic-plastic deformation impacts diffusion through the stress field, governing the pressure dependence of the bulk chemical potential. Second, mass transport affects the fracture resistance via H buildup in the fracture process zone, reducing the critical energy release rate. And third, the H-sensitive phase field degrades the strain energy density of the solid.\\

We solve the linearized finite element system,
\begin{equation}
\begin{bmatrix}
  \bm{K}^u & 0 & 0\\
  0 & \bm{K}^\phi & 0 \\
  0 & 0 & \bm{K}^C 
 \end{bmatrix} \begin{bmatrix} \bm{u} \\ \bm{\phi} \\ \bm{C} \end{bmatrix} + 
 \begin{bmatrix}
  0 & 0 & 0\\
  0 & 0 & 0 \\
  0 & 0 & \bm{M}
 \end{bmatrix} \begin{bmatrix} \dot{\bm{u}} \\ \dot{\bm{\phi}} \\ \dot{\bm{C}} \end{bmatrix}= \begin{bmatrix} \bm{r}^u \\ \bm{r}^\phi \\ \bm{r}^C \end{bmatrix}
\end{equation}

\noindent by means of a time parametrization and an incremental-iterative scheme in conjunction with the Newton-Raphson method. A time increment sensitivity analysis is conducted in all computations. The modeling framework is implemented in the commercial finite element package ABAQUS via a user element (UEL) subroutine. Post-processing of the results is carried out by means of Abaqus2Matlab \cite{AES2017}.


\section{Propagating hydrogen boundary conditions}
\label{App:MovingBCs}

In the presence of a propagating crack, the appropriate chemical boundary conditions on the newly exposed surface require careful consideration. As discussed in Section \ref{Sec:Results}, capturing how H transport is driven through a propagating crack could be key in understanding H embrittlement under in-situ charging. A suitable approach is to assume that the environment will promptly occupy the space created with crack advance. Accordingly, the value of the H concentration corresponding to the choice of electrochemical solution and applied potential, $C_{env}$, should be prescribed on the new boundary emerging due to crack propagation. Such scheme could be extended to account for chemical reaction and mass transport limitations \cite{Turnbull2015}.\\

In this work, we choose to effectively prescribe the H concentration in the damaged regions by means of a penalty approach \cite{Renard2019}. An alternative procedure is to define a moving boundary via general multipoint constraints (MPCs), see Ref. \cite{EFM2017}. We choose to enforce $C=C_{env}$ in the cracked domain ($\phi=1$), ramping linearly the H concentration from the $\phi>0.5$ regions by choosing a sufficiently large value of the penalty factor $k_p$. Thus, we add an extra term to the chemical residual Eq. (\ref{eq:ResH}), which now reads
\begin{equation}
       r_{i}^C = \int_\Omega \left\{ N_i^T \left[ \frac{1}{D} \frac{dC}{dt} + \left(  C - C_{env} \right) \langle 2 \phi - 1 \rangle k_p \right] + \bm{B}_i^T  \nabla C V  - \bm{B}_i^T \left( \frac{\bar{V}_H C}{RT}  \nabla \sigma_H \right)  \right\} \, \mathrm{d}V
\end{equation}

\noindent where $\langle \, \rangle$ denote the Macaulay brackets, and $C$ and $C_{env}$ are integration point quantities. Accordingly, the diffusivity matrix Eq. (\ref{eq:KH}) becomes
\begin{equation}
    \bm{K}_{ij}^C = \int_\Omega \left(\bm{N}_i^T  \bm{N}_j \langle 2 \phi - 1 \rangle k_p  +\bm{B}_i^T  \bm{B}_j  - \bm{B}_i^T \frac{\bar{V}_H}{RT}   \nabla \sigma_H N_j   \right) \, \mathrm{d}V    
\end{equation}

\begin{figure}[H]
\centering
\includegraphics[scale=0.8]{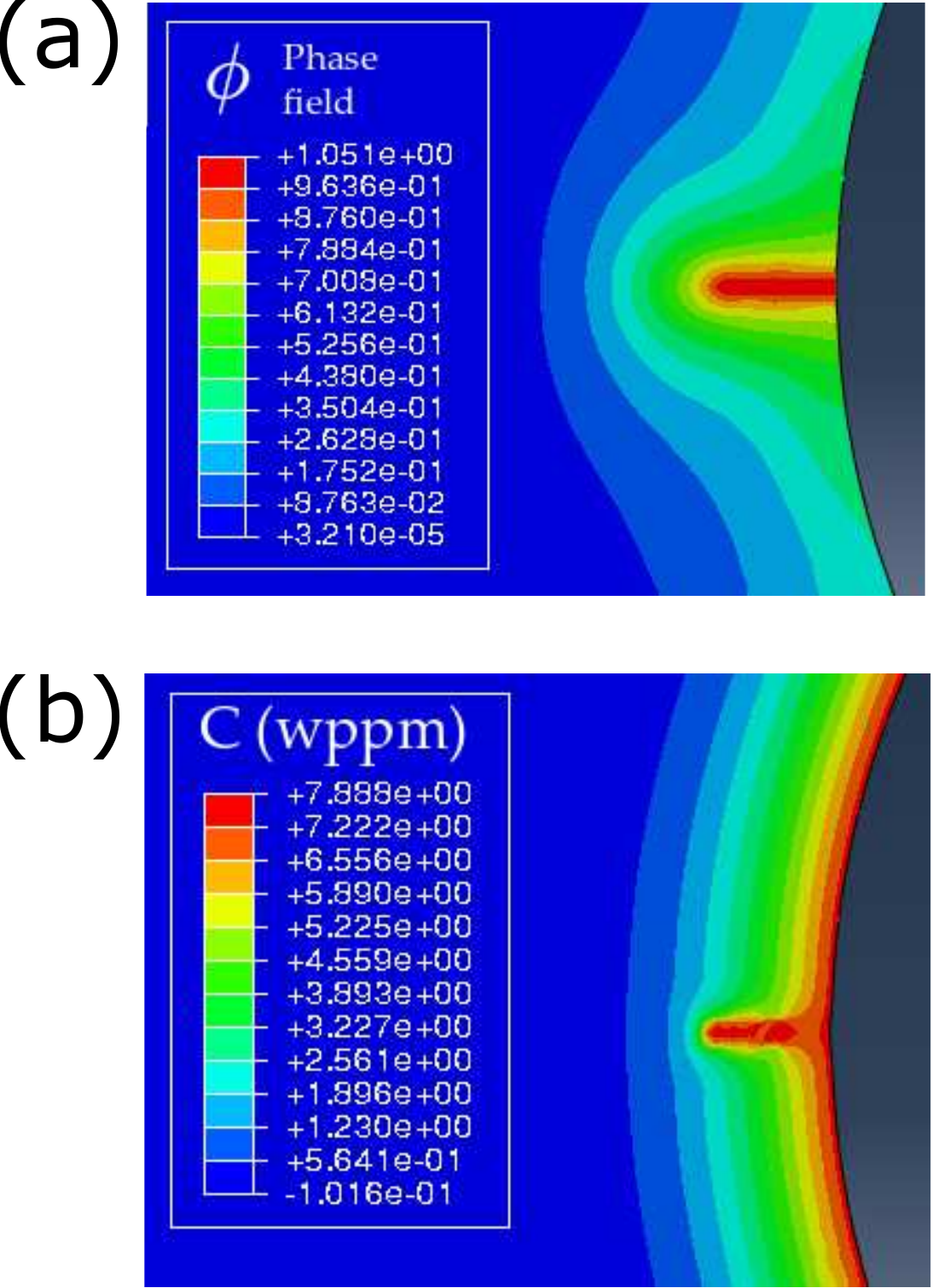}
\caption{Penalty boundary conditions for hydrogen diffusion at the newly created surfaces. Representative results of hydrogen propagation with crack advance, (a) phase field fracture parameter, and (b) hydrogen concentration.}
\label{fig:MovingBCs}
\end{figure}

Values of $k_p$ as high as $1 \times 10^6$ effectively ensure $C=C_{env}$ in the damaged regions without hindering convergence. Representative contours are shown in Fig. \ref{fig:MovingBCs} for the Allvac lot under an applied potential of $E_A=-1100$ mV ($C_{env}=7.54$ wppm, see Table \ref{Tab:DiffusibleH}). 

\bibliographystyle{elsarticle-num}
\bibliography{library}

\end{document}